# Reactive Laser Synthesis of Ultra-high-temperature ceramics HfC, ZrC, TiC, HfN, ZrN, and TiN for Additive Manufacturing

Adam B. Peters[1], Chuhong Wang[1], Dajie Zhang[1,2], Alberto Hernandez[1], Dennis C. Nagle[1,2], Tim Mueller[1], James B. Spicer[1,2]*

[1] Department of Materials Science and Engineering, The Johns Hopkins University

3400 North Charles Street, Baltimore, MD 21218

[2] The Johns Hopkins Applied Physics Laboratory, Research and Exploratory Development Department, 11100 Johns Hopkins Road, Laurel, MD 20723

## Graphical Abstract

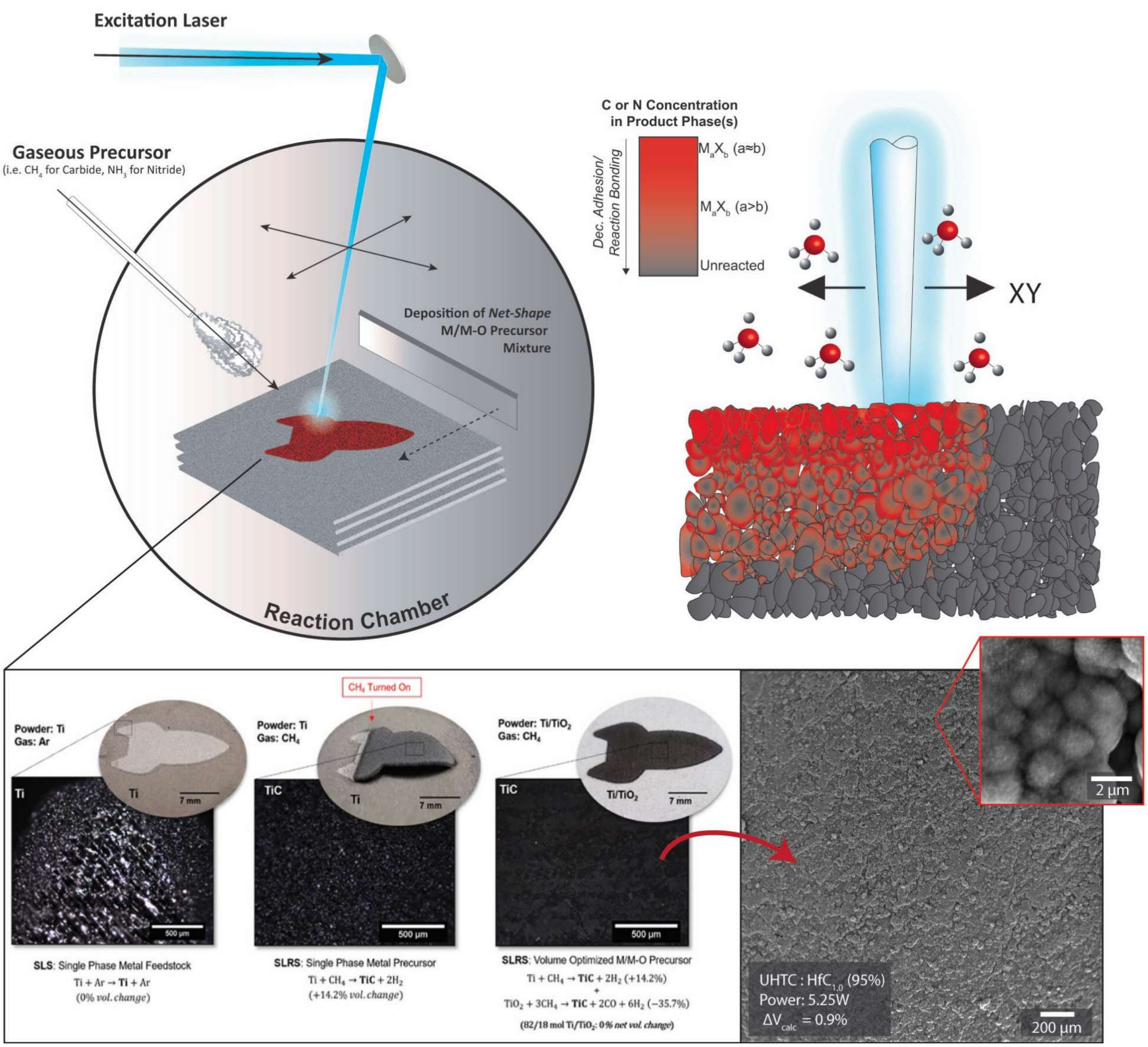

## Abstract

Ultra-high-temperature ceramics (UHTCs) are optimal structural materials for applications that require extreme high temperature resilience ($M_p$>3000 °C), resistance to chemically aggressive environments, wear, and mechanical stress. Processing UHTCs with laser-based additive manufacturing (AM) has not been fully realized due to a variety of obstacles. In this work, selective laser reaction sintering techniques (SLRS) were investigated for the production near net-shape UHTC ceramics such as HfC, ZrC, TiC, HfN, ZrN, and TiN. Specifically, group IV transition metal and metal oxide precursor materials (<44 μm) were chemically converted and reaction-bonded into layers of UHTCs using single-step selective laser processing in 100 vol% $CH_4$ or $NH_3$ gas that might be compatible with prevailing powder bed fusion techniques. Conversion of either metals (Hf, Zr and Ti) or metal oxides ($HfO_2$, $ZrO_2$, and $TiO_2$) particles was first investigated to examine reaction mechanisms and volume changes associated with SLRS of single-component precursor systems. SLRS processing of metal or metal oxide alone produced near stoichiometric UHTC phases and yields up to >99.9 wt% total for carbides and nitrides and during the rapid reactive *in-situ* processing scheme. However, for single-phase feedstocks, gas-solid reactivity induced volumetric changes (correlated with the stochiometry of the rocksalt-type UHTC carbide and nitride products) resulted in residual stresses and cracking in the model-AM product layer. To mitigate conversion-induced stresses of single-phase precursors, composite metal/metal oxide precursors were employed to compensate for the volume changes of either the metal (which expands during conversion) or the metal oxide precursor (which contracts). While conversion of the optimized composite materials produced HfC layers with as little as +0.9% volume change, results indicated interparticle adhesion must be optimized to obtain robust UHTC-AM layers. Computational models of carbon and nitrogen diffusion in host transition metal lattices corroborated experimental results where a progressive particle conversion might inhibit interparticle diffusion unless laser processing parameters are carefully optimized to favor reaction bonding over discrete particle conversion. While this method presents a host of processing considerations, we demonstrate how this reactive approach may be viable for the AM of numerous UHTC materials that are not readily produced using current methods.

# 1 Introduction

Ultra-high-temperature-ceramics (UHTCs) and their composites are typically defined as non-metallic, inorganic solids with melting temperatures exceeding 3000 °C and/or thermal and chemical stability in air above 2000 °C. While several functional definitions exist, UHTCs are chemically characterized as binary compounds comprised of a metal atom (typically an early transition metal such as Ti, Zr, Hf, *etc.*) that is bonded to a non-metal such as C, N, or B [1]. UHTCs encompass a large set of non-oxide materials and maintain a unique combination of metallic- and ceramic-like properties [2]. The strong covalent bonds in these transition metal carbide, nitride and boride materials produce highly desirable properties: extremely high melting temperatures-temperatures; high thermal and electrical conductivity; and resistance to chemically reactive conditions, radiation, stress, and mechanical wear. As a result, recent interest in UHTCs has been catalyzed by the limitations of traditional refractory materials for aerospace applications, rocket propulsion, and hypersonics [3]. In particular, Group IV, UHTC carbides, and nitrides have received attention for thermal protection systems, nozzle throats, and control thrusters which must serve under a combination of high thermal and mechanical loads, aggressive oxidizing environments, and rapid heating rates sustained during flights beyond Mach 5. [1]–[5]. The characteristic properties of Group IV UHTCs extents to other extreme environment applications; a summary of the properties and applications are provided in Table 1 [3], [6]–[12].

**Table 1. Properties and Applications of Important UHTCs Carbides and Nitrides Selected for SLRS Processing** [13][14].

| Target UHTC | Melting Point (K) | Young's Modulus (GPA) | Thermal Conductivity ($W\ m^{-1}\ K^{-1}$) | Applications |
|---|---|---|---|---|
| **HfC** | 4228 | 420 | 25.1 | Rocket combustion chambers, arc-jet thrusters, thermal protection systems, rocket nozzles, solar energy applications, oxidation coating for C/C composites and superalloys [7], [9] |
| **ZrC** | 3718 | 350 | 24.6 | Nuclear barrier coatings, field emitters, integrated circuits [15] |
| **TiC** | 3365 | 450 | 28.9 | Heat shields, cutting tools, wear-resistant components, [7] |
| **HfN** | 3628 | 380 | 11 | Rocket nozzles, solar energy applications, plasmonic materials, wear-resistant coatings [3] [10] |
| **ZrN** | 3298 | 460 | 11 | Matrix candidate for nuclear fuel [8], solar energy applications, plasmonic materials [10], dental and surgical implants [11], fuel cell membrane [9] |
| **TiN** | 3348 | 420 | 29 | Diffusion barriers, crucibles for radioactive metals, wear-resistance coatings [3] |

Processing refractory ceramics into complex geometries is challenging and costly. Most UHTC's are synthesized by carbothermal reduction of transition metal oxides before being processed as dry powders or colloidal shaping techniques [2]. Even with advanced ceramic processing techniques, the consolidation of UHTC particles is difficult due to strong atomic bonding and low diffusion characteristics. Competition between densifying and coarsening mechanisms during processing makes high densities not easily obtainable even with pressure-assisted techniques (*i.e.* spark plasma sintering or hot isostatic pressing). Geometric complexity is often limited to two-dimensional plasma sprayed coatings on metallic substrates or to simple axially-symmetric shapes (like cylinders or tiles) when sufficient densification is achieved [3]. As a result, UHTC shapes cannot often satisfy applications that require complicated part geometries that might be otherwise produced using additive manufacturing (AM) [16], [17].

Conventional AM methods (for both refractory metals and lower melting temperature ceramics) are not easily adapted to refractory non-oxide ceramics. Generally, additive manufacturing of refractory ceramics requires binder phases or organic additives (dispersants, binders, plasticizers, lubricants, etc.) to shape non-reactive ceramic feedstocks and impart desired rheological properties. High-temperature sintering (> 2000°C) is then required after shaping, however, covalent-ionic bonds inhibit sufficient atomic mobility to relieve thermally-induced stresses so defects commonly prevent appreciable mechanical integrity from being obtained [18], [19], [20]. In contrast to these indirect AM methodologies, single-step direct AM approaches like selective laser sintering (SLS) and selective laser melting (SLM) can lead to decomposition or microcracking when these materials are heated to temperatures that produce mobility. Consequently, the ability to additively manufacture the next generation of robust components for extreme high-temperature environments has not been extensively explored.

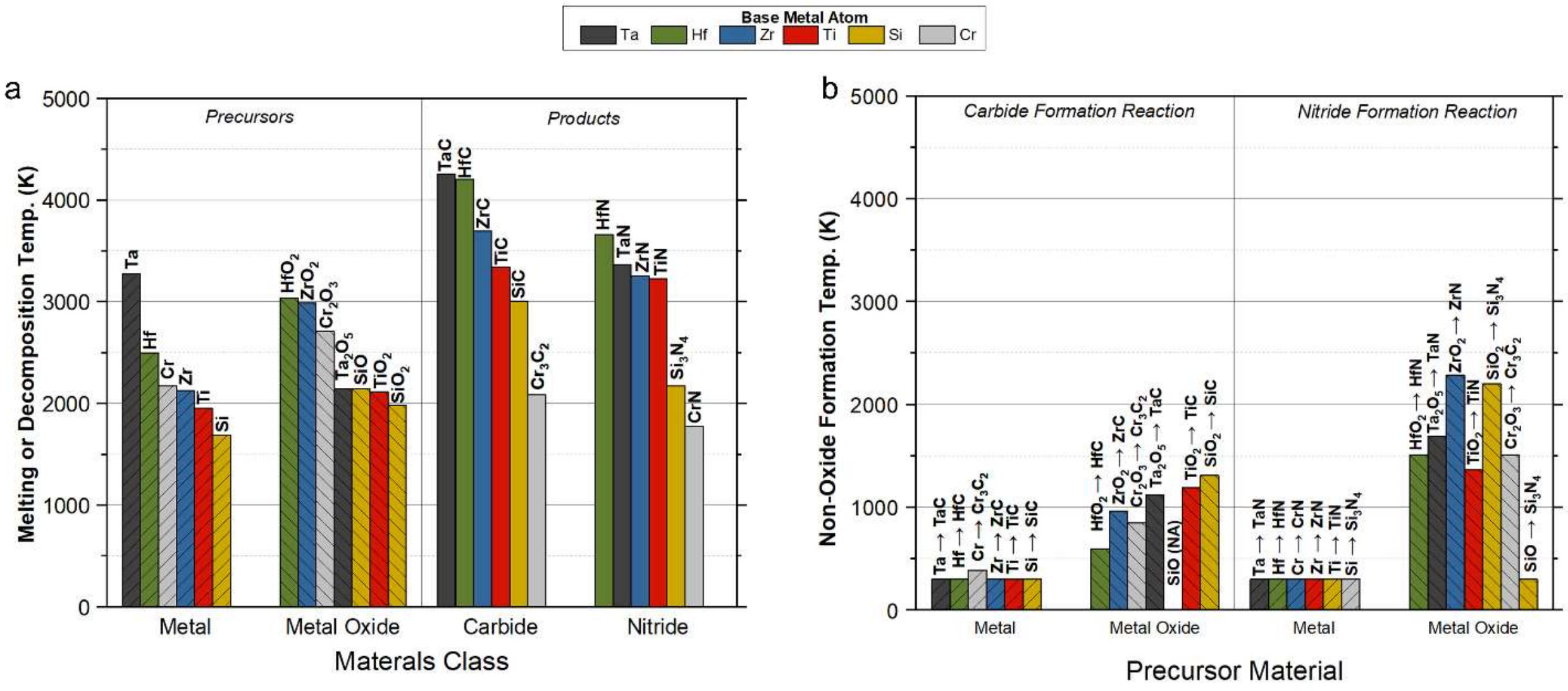

**Figure 1.** (a) Melting (or decomposition temperatures) of carbide and nitride ceramics produced using SLRS processing compared to the estimated spontaneous synthesis temperatures (b) of their corresponding metal or metal oxide precursors in $CH_4$ for carbides and $NH_3$ for nitrides. The SLRS synthesis $Cr_3C_2$ and CrN were previously demonstrated as a model system in [21].

Because UHTCs powders are readily synthesized from metal oxides at low temperatures, an alternative UHTC-AM approach is to combine simultaneous carbothermal reduction (or nitridation) with selective laser processing using SLRS [21] (illustrated in Fig. 2). Laser gas nitridation or carburization for surface treatment of metals has been the subject of significant investigation, yet the *in-situ* synthesis of UHTCs using SLRS-like processing has not been widely demonstrated beyond a handful of studies [22], [23], [24]. Laser-induced synthesis has typically been limited to either the nitridation of bulk metals or nanoparticles in gas/liquid phase $N_2$ ([23], [25]–[27]) or by carburization using solid carbonaceous phases [9],[28]. These techniques have not been applied to the development of powder bed fusion additive manufacturing.

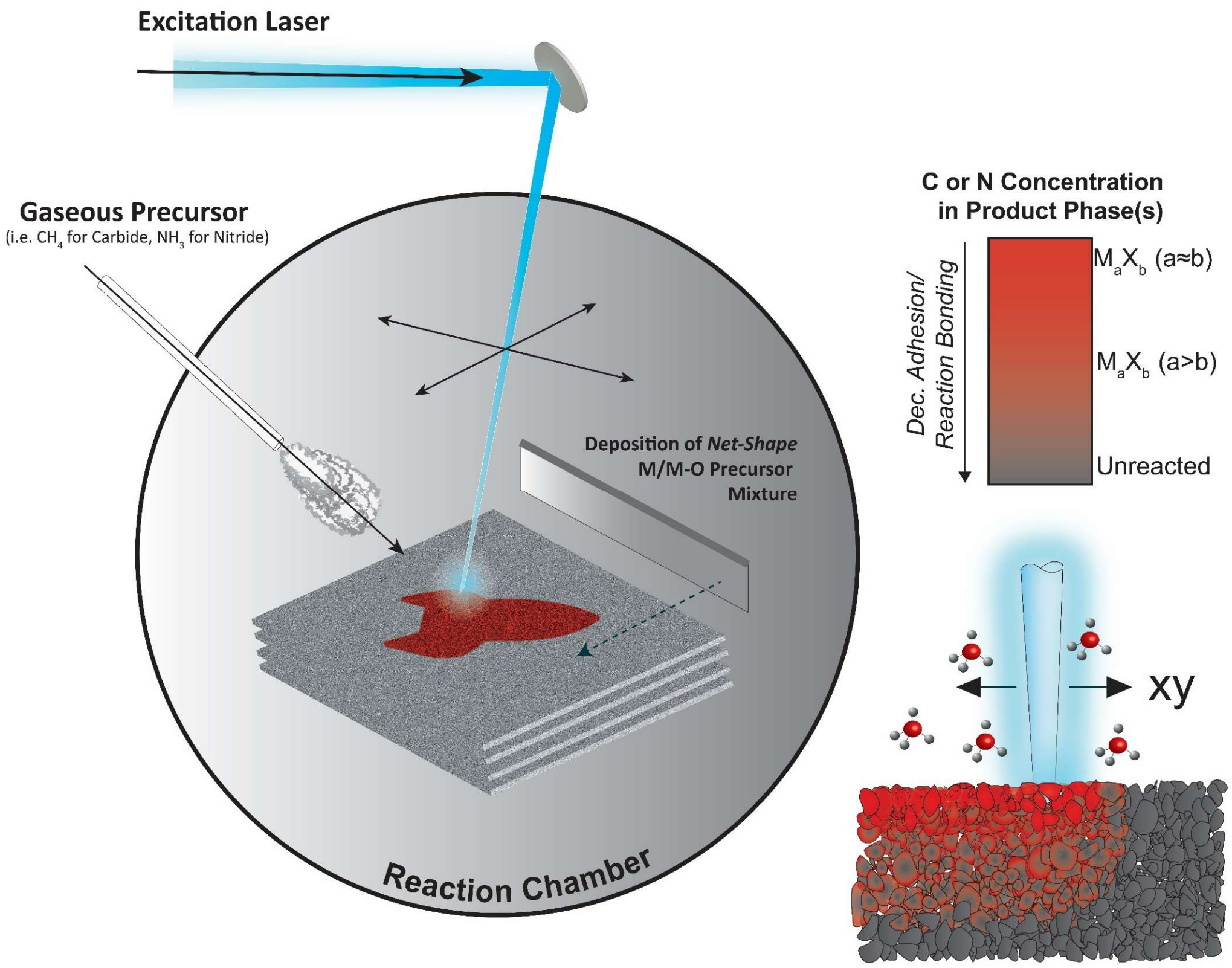

**Figure 2**. SLRS processing scheme depicting a reaction chamber containing metal and or metal oxide precursor materials that are laser irradiated within the presence of a gaseous precursor ($CH_4$ or $NH_3$) (Left). Photothermochemical gas-solid reaction results in the simultaneous synthesis and reaction bonding of a precursor later [21]. Additional layers may then be deposited over the reaction bonded later and the process may be repeated. Schematic overview of the selective laser reaction sintering processes depicting the thermal zones in relation to the depth-dependent conversion of the powder bed; and the propagation of heterogeneous depth-dependent conversion throughout the precursor matrix (Right). In the color-coded legend, $M_aX_b$ can represent either a non-oxide phase. When $a \approx b$ the phase is carbon or nitrogen-rich, when a>b the phase may be carbon or nitrogen deficient.

Our previous work demonstrated the promising viability of SLRS for reactive non-oxide AM with the *in-situ* synthesis of near $Cr_3C_2$ and CrN refractory ceramics from $Cr/Cr_2O_3$ precursors in $CH_4$ and $NH_3$ atmospheres [21], [29], [30]. Here, the conversion of the metal precursor layer resulted in specific volume increases that led to layer buckling, while the conversion of the metal oxide precursor caused a volume reduction that led to a mudcrack-like surface microstructure [21]. Although precursor particle morphology was found to influence the homogeneity of the final product layer, conversion-induced volume changes were observed to occur independently of particle size and reactant gas. To create ~~a~~ crack-free, near net-shape carbide or nitride product layers, the ratio of metal and metal oxide precursor constituents were optimized to prevent the accumulation of residual stresses that accompanied *in-situ* conversion. This work demonstrated several benefits of the SLRS approach [29], [30]:

(1) High-temperature refractory materials can be created through reaction bonding with relatively low laser energy to circumvent issues associated with direct laser melting and sintering [31];
(2) Carbide and nitride ceramics can be synthesized using laser irradiation in gaseous atmospheres at temperatures below traditional sintering temperatures by selecting appropriate precursors;

(3) Net-shape (isovolumetric) may be obtained when the ratio of reactive precursor constituents (metals and metal oxides) are optimized to compensate for the respective volume expansions and contractions of the metal and metal oxide precursor materials upon conversion [28], [30] using the following equation:

$$f_{\Delta V=0} = \frac{v_{mo} - y v_{mx}}{v_{mo} + (1-y) v_{mx} - v_m}, \quad \text{(Eq. 1)}$$

where $f$ is the mole fraction of metal in the composite precursor for zero volume change, $v_{mx}$ is the molar volume of the metal carbide or nitride, $v_m$ is the molar volume of the metal, $v_{mo}$ is the molar volume of the oxide, and $y$ is the oxidation state of M in the product (i.e +2) [29].

This same approach may be broadly translated to the *in-situ* processing of other refractory materials beyond the $Cr_3C_2$ and CrN model systems and to ultra-high temperature carbides and nitrides under the appropriate reaction conditions.

The aims of the work described here are to:

(1) Systematically investigate the chemical conversion of transition metal and metal oxide materials to different UHTC materials systems using techniques that are compatible with SLRS-AM;
(2) Examine the reaction mechanism and volume changes of single-component metal or metal oxide systems in the context of precursor morphology;
(3) Assess associated volumetric changes to perform near net-shape synthesis that might be applied to additive manufacturing.

Several of these materials have not been produced using laser reaction-synthesis techniques. In this work, a combination of computational and experimental techniques were used to investigate the thermodynamic and kinetic phenomena that govern the conversion and reaction bonding of precursor materials to UHTCs. In Tables 2 and 3, the reactions of solid precursors with $CH_4$ and $NH_3$ for UHTC carbides and nitride are shown.

**Table 2. Isovolumetric Formulations of Carbide Ultra-High Temperature Ceramics Using Gas-Solid Reactions.**

| Product ($M_p$) | Solid Phase Precursor | Solid Precursor Mp (°C) | Reaction | Reaction with $CH_4$ | Est. Spontaneous Reaction Temp. (°C) | Volume Change (%) | M/M-O Ratio for Net-Shape Conversion in Mol. (wt%) |
|---|---|---|---|---|---|---|---|
| **TiC** | Ti | 1668 | 1 | $Ti + CH_4 \rightarrow \mathbf{TiC} + 2H_2$ | <25 | +14.2% | 82/18 (73/17 wt%) |
| (3092°C) | $TiO_2$ | 1843 | 2 | $TiO_2 + 3CH_4 \rightarrow \mathbf{TiC} + 2CO + 6H_2$ | 913 | -35.7% | 82/18 (73/17 wt%) |
| **ZrC** | Zr | 1855 | 3 | $Zr + CH_4 \rightarrow \mathbf{ZrC} + 2H_2$ | <25 | +11.2% | 79/21 (74/26 wt%) |
| (3445°C) | $ZrO_2$ | 2715 | 4 | $ZrO_2 + 3CH_4 \rightarrow \mathbf{ZrC} + 2CO + 6H_2$ | 685 | -27.8% | 79/21 (74/26 wt%) |
| **HfC** | Hf | 2233 | 5 | $Hf + CH_4 \rightarrow \mathbf{HfC} + 2H_2$ | <25 | +14.3% | 74/26 (70/30 wt%) |
| (3995°C) | $HfO_2$ | 2758 | 6 | $HfO_2 + 3CH_4 \rightarrow \mathbf{HfC} + 2CO + 6H_2$ | 319 | -28.1% | 74/26 (70/30 wt%) |

*values calculated based on material density changes

**Table 3 Isovolumetric Formulations of Nitride Ultra-High Temperature Ceramics Using Gas-Solid Reactions.**

| Product | Solid Phase Precursor | Solid Precursor Mp (°C) | Reaction | Reaction with $NH_3$ | Est. Spontaneous Reaction Temp. (°C) | Volume Change (%) | M/M-O Molar Ratio for Net-Shape Conversion Mol. (wt%) |
|---|---|---|---|---|---|---|---|
| **TiN** | Ti | 1668 | 9 | $2Ti + 2NH_3 \rightarrow \mathbf{2TiN} + 3H_2$ | <25 | +11.6 | 85/15 (77/23 wt%) |
| (2930°C) | $TiO_2$ | 1843 | 10 | $6TiO_2 + 4NH_3 \rightarrow \mathbf{6TiN} + 12H_2O + N_2$ | 1088 | -37.2 | 85/15 (77/23 wt%) |
| **ZrN** | Zr | 1855 | 11 | $2Zr + 2NH_3 \rightarrow \mathbf{2ZrN} + 3H_2$ | <25 | +5.6 | 90/10 (87/13 wt%) |
| (2980°C) | $ZrO_2$ | 2715 | 12 | $6ZrO_2 + 4NH_3 \rightarrow \mathbf{6ZrN} + 12H_2O + N_2$ | 2008 | -31.6 | 90/10 (87/13 wt%) |
| **HfN** | Hf | 2233 | 13 | $2Hf + 2NH_3 \rightarrow \mathbf{2HfN} + 3H_2$ | <25 | +2.9 | 95/5 (94/6 wt%) |
| (3385°C) | $HfO_2$ | 2758 | 14 | $6Hf + 4NH_3 \rightarrow \mathbf{6HfN} + 12H_2O + N_2$ | 1235 | -35.4 | 95/5 (94/6 wt%) |

*values calculated based on material density changes

# 2 Methods

## 2.1 Precursor selection and System Preparation

Various refractory metal and metal oxide phases were selected for gas-solid conversion to carbide and nitride UHTCs by reaction with $CH_4$ and $NH_3$ gas respectively. Precusor materials <44 μm (-325 mesh) were selected for trial SLRS conversion due to their relative size similarity to SLM and SLS feedstocks which range from 10-50 μm in diameter [9], [13]. The average particle size of commercially-available precursor materials ranged from <1 - 40 μm despite having the same upper limit mesh size. For the initial processing, neither modification of particle size nor alteration of morphology was conducted. . The literature and our previous SLRS research indicate that small precursor particles (<10 μm) facilitate conversion due to higher specific surface areas, but precursors composed of only small particles impose challenges that have been noted [21], [28], [30]. By contrast, precursors containing larger particles were not as readily converted but produced more favorable, crack-free SLRS products. SEM images of precursors are presented in Fig. 3 while the characteristics of the precursor materials used for SLRS conversion to carbide and nitride UHTCs are provided in Table 4.

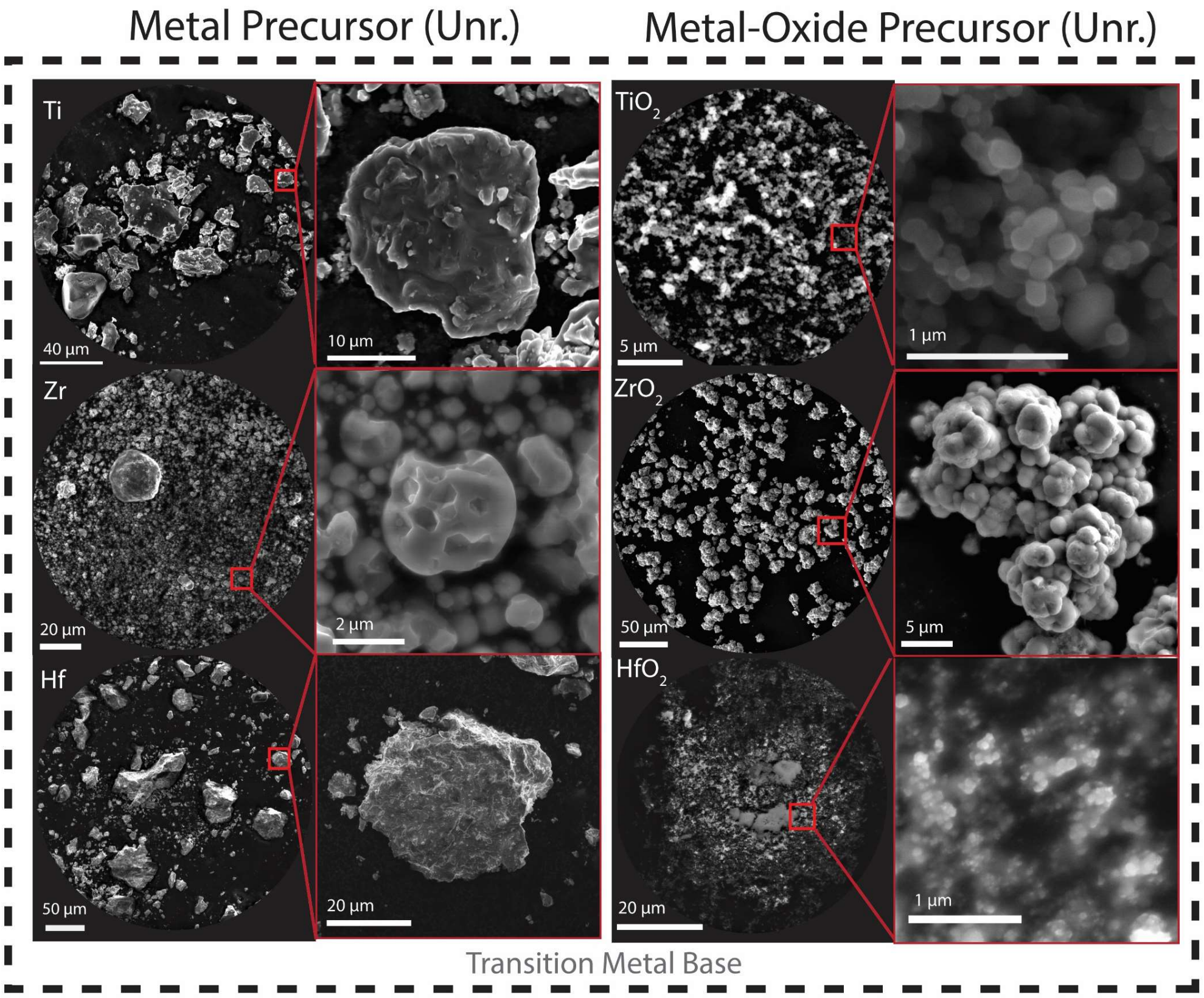

**Figure 3.** SEM images of metal and metal-oxide precursor materials used for SLRS processing. The lower magnification image on the right illustrates the variation in particle size for a given precursor, while the image on the left show typical particle morphology. All precursor materials were commercially available (<44um, 325 mesh sizes), however, the average particle size varied.

Of the precursors, selected in this work Ti, Hf, and $ZrO_2$ had approximate particle sizes (APS) in the 10-50 µm PBF range. By contrast, the APS of $TiO_2$, $HfO_2$, and Zr had considerably smaller sizes.

**Table 4. Selected Precursor Powder Characteristics***

| Name | Feedstock Powder | Chemical Purity (wt. %) | K-M, Absorbance (a.u.) at 445 nm | Packing Density | Particle Morphology |
|---|---|---|---|---|---|
| Ti | Alfa Aesar | 99% | 0.47 | 0.44 | Rough, Dense |
| $TiO_2$ | Cerao | 99.9% | $1.6\times10^{-3}$ | 0.40 | Round, Sub-Micron, Dense |
| Zr | Alfa Aesar | 99% | 0.61 | 0.57 | Round, Smooth |
| $ZrO_2$ | Cerao | 99% | $3.3\times10^{-3}$ | 0.43 | Agglomerated, Porous |
| Hf | Atl. Equip. Engineers. | 99.5% | 0.52 | 0.50 | Rough, Blocky, Dense |
| $HfO_2$ | Alfa Aesar | 99% | $2.6\times10^{-3}$ | 0.25 | Round, Sub-Micron, Dense |

*all particles were listed as "-325 mesh" (<44 µm)

## 2.2 Phase and Microstructural Characterization

Metal and metal oxide precursors laser-processed in $CH_4$ or $NH_3$ were characterized using x-ray diffraction (XRD) to determine the rate of conversion to carbide and nitride phases. Due to the fragility of the converted single-layer samples, XRD characterization was performed with the UHTC materials still supported on their precursor powder beds. Quantitative phase characterization was performed from 20° to 80° 2θ using Rietveld refinement with the aid of Match! (version 3) software. When synthesized via carbothermal or nitridation reduction of oxides, non-oxide products are known to contain some level of oxygen impurities [8], [9], [32]. All group IV transition metals carbides and nitrides form continuous solid solutions in the rock salt structure with oxygen impurities. For SLRS reaction products containing meaningful oxygen concentrations in the NaCl-type crystal structure, lattice parameters of $M_xC_yO_z$ (or $MC_yO_{(1-y)}$, where z= 1-y) the ratio of C (or N) to O dictates the lattice parameter and can be calculated from $MC_y$ and $MO_z$ according to Vegard's law and similar empirically-derived relationships [33], [34]:

$$a = (1 - y)a_{MO} + ya_{MC} \ , \qquad \text{(Eq. 2)}$$

where *a* is the lattice parameter and *x* is the fraction of the metal oxide (*MO*) or metal carbide/nitride (*MC*) phase. Lattice parameters of non-stoichiometric carbides, nitrides oxycarbides, and oxynitrides were estimated from lattice parameter-stoichiometry literature data and application of Vegard's law where applicable. Lattice parameters obtained from Rietveld refinement were used to calculate the net volume change metal/metal-oxide precursor mixtures. Vacancy-related effects were not incorporated into the estimations of lattice parameters for oxycarbides and oxynitride species due to a lack of data available for the estimation of O/C/N/vacancy relationships.

Both optical and SEM microscopy were employed to study the morphologies and microstructures of converted precursors to determine how particle size and processing parameters affected the resultant structures. Additional details regarding characterization can be found elsewhere and the methods used in this work are identical to those previously described [21].

## 2.3 Kinetic Modeling of Inter-Lattice Diffusion

Variations in interstitial occupancy have a profound effect on inter-lattice diffusion of C or N and impact the rates of gas-solid conversion. Compared to the narrow range of stoichiometries for intermediate transition metal carbides and nitrides [21] (e.g. Cr-C system that forms, non-FCC or BCC phases like $Cr_3C_2$, $Cr_7C_3$, and $Cr_{23}C_6$ with narrow stoichiometric ranges (~3 at%) [35]), the carbides and nitrides of Ti, Zr, and Hf form unique NaCl-type crystal structures with a wide range of stoichiometries up to ~ 50 at%. For example, group IV carbide and phases are very similar: monocarbides $TiC_{1-x}$ $ZrC_{1-x.}$ $HfC_{1-x}$, all show extensive homogeneity ranges and high congruent melting temperatures. Meanwhile, the analogous nitride phases, while less studied, are generally multi-phasic. For example, the Ti-N system contains a high solubility of nitrogen in α-Ti with three sub-nitride phases, $Ti_3N_{2-x}$, $Ti_4N_{3-x}$, $Ti_2N$, and $TiN_{1-x}$. Kinetic modeling for inter-lattice diffusion was conducted using *ab initio* calculations and machine-learned interatomic potentials in molecular dynamics simulations to provide a fundamental framework for underlying reaction mechanisms associated with experiments.

### *2.3.1 Ab initio molecular dynamics (AIMD) and Machine learning interatomic potential driven molecular dynamics*

The ionic conductivities of phases in the Ti-N-C chemical space were computed using a scheme that was developed recently [36] and is based on learning-on-the-fly molecular dynamics (LOTF-MD) and moment tensor potentials (MTP) [37]. The use of this scheme enables the generation of orders of magnitude more data than can be achieved using AIMD alone with little loss of accuracy in the calculated energies, resulting in more accurate estimates of ionic conductivity. These calculations were performed using a recently-developed scheme [36] based on learning-on-the-fly molecular dynamics (LOTF-MD) and moment tensor potentials (MTP) [38]. The use of this scheme enables the generation of orders of magnitude more data than can be achieved using only density functional theory with little loss of accuracy in the calculated energies, resulting in more accurate estimates of ionic conductivity.

The moment tensor potential for each phase was trained using the Machine Learning of Interatomic Potentials (MLIP) software package [39]. The training data for the initial potential was extracted with 10-fs time intervals from a 13 ps molecular dynamics simulation calculated using density functional theory. Molecular dynamics simulations used to evaluate ionic conductivity were performed in LAMMPS [40]. The interface between MTP molecular dynamics and DFT static calculations was also carried out by MLIP. Details of the active learning criteria, as well as experimental benchmark results, can be found in a previous publication [36]. Multi-temperature molecular dynamics simulations for each phase were performed to determine the activation energy for diffusion by fitting an Arrhenius relationship. The simulation time was at least 10 ns for each phase.

## 2.4 Experimental Design

UHTC precursors were fashioned into powder beds representative of SLS and SLM-AM techniques using a doctor blade screening method. Powders were poured into cylindrical reservoirs (stainless steel, inner diameter: 25.5 mm, depth: 4.5 mm) and a metal blade was used to screed each precursor bed into an even layer and create a flush powder bed surface. Each precursor bed was then inserted into the laboratory-scale SLRS reaction chamber for selective laser processing.

The reaction chamber consisted of a quartz tube chamber (46 x 50 x 310 mm), LabVIEW-controlled gas flow regulators, and a one-way gas exhaust. A 445 nm, 4 W or 8 W (optical power output) transistor-transistor-logic diode laser with an integrated focusing lens was utilized for laser heating. The diode laser was mounted to a two-axis XY linear scanning system. The 445 nm wavelength was ideal to minimize gas-phase absorption by $CH_4$ and $NH_3$ that occurs in LWIR (e.g. at 10.6 μm) to prevent unstable power fluctuations observed by others [41]. The 4 W, 445nm diode laser was set to 3 W and 4 W power settings with a translational scan speed of 100 mm/min in initial conversions of single-component metal/metal oxide precursors to study base reaction thresholds. The 4 W, 445 nm diode laser produced a power density of approximately 8 W/mm$^2$ (3 W, beam diameter ~750 μm) and 32 W/mm$^2$ (4 W, 400 μm). Different laser optics were used to refine the spot size and increase energy density. The power density was calculated using $1/e^2$ gaussian profiles for the 3 W and 4 W optical output beams). Although the beam was non-uniform, power densities were estimated to be 14 W/mm$^2$ and 64 W/mm$^2$ for the 3 W and 4W beams respectively. In several cases where no or incomplete SLRS reaction was observed, energy density was increased from 3 W or 4 W to 7 W to facilitate reactivity after initial trials. Laser powers of 4-7 W were used for net-shape composite precursor experiments to aid the conversion of precursor materials that underwent incomplete carbide or nitride formation [21]. The 7 W power setting of the diode laser was not able to be effectively used for carbide conversion, as low levels of optical absorption by the quartz tube caused quartz heating and spontaneous carbon deposition on the interior surface of the reaction chamber. This effect prevented laser penetration through the quartz window and onto the precursor bed. Solid decomposition products from $NH_3$-containing atmospheres were not observed so power could be more easily increased to the highest output power achievable using the laser diode. All scan parameters are summarized in Table 5. Lightburn software was used to control integrated power, scan rate, and beam path. The scan rates and optical output used in this study are significantly lower than those used in standard AM machines (~100 - 1500 mm/s [42], [43] for inert processing). However, for other non-additive processes such as laser nitridation of metals, the scan speeds are comparable to those used here [44].

**Table 5. SLRS Laser Processing Parameters**

| | Low Power | Medium Power | High Power | Max Power |
|---|---|---|---|---|
| Wavelength | (λ) = 445nm | (λ)= 445nm | (λ)= 445 nm | (λ)= 445nm |
| Average Power | (P) = 3.0 W | (P) = 4.0 W | (P) = 5.25 W | (P) = 7 W |
| Spot Diameter. | (D) = ~400 μm | (D) = ~750 μm | (D) = ~300 μm | (D) = ~300 μm |
| Power Density* | (F) = 8W/mm$^2$ | (F) = 32W/mm$^2$ | (F) = 58W/mm$^2$ | (F) = 78W/mm$^2$ |
| Processing Speed | (V) =100 mm/min | (V) = 100 mm/min | (V) = 100 mm/min | (V) = 100 mm/min |
| Processing Gas | 100 vol.% $NH_3$ or $CH_4$ | 100 vol.% $NH_3$ or $CH_4$ | 100 vol.% $NH_3$ or $CH_4$ | 100 vol.% $NH_3$ or $CH_4$ |
| Gas Flow rate | 250 cm$^3$/min | 250 cm$^3$/min | 500 cm$^3$/min | 500 cm$^3$/min |

*Note, some transient optical output fluctuations (+/- 10%) were observed during processing.

## 2.5 Experimental Process

After each precursor powder bed was introduced into the reaction chamber, the chamber was purged with high-purity argon (500 SCCM) to prevent oxidation during laser heating. The reaction chamber was filled with flowing $CH_4$ or $NH_3$ (250 or 500 SCCM) for carbide and nitride synthesis. The faster flow rates of 500 SCCM were used for higher-power laser studies to minimize the deposition of vaporized precursor materials (especially for small particles) on the quartz tube wall and provide greater gas-phase availability. With higher laser energy densities, oxide particles can be vaporized into the chamber. Faster gas flow helped remove vaporized reaction products and prevent absorptive residue from readily adhering to the quartz reaction tube.

Although stable at room temperature, ammonia is unstable above 459 K and is efficiently thermally activated at ~823 K [45]. Ammonia readily dissociates into gaseous nitrogen and hydrogen according to the chemical equilibrium at the surface of metal via catalytic reactions (Eq. 3) [46].

$$NH_3 \rightleftharpoons N_{(ad)} + 3/2\, H_2 \qquad \text{(Eq. 3)}$$

The successive dehydrogenation steps for the ammonia molecule lead to various adsorbed radicals on the surface, where through absorption, diffusion, and reactions, metal nitrides form with decreasing nitride content according to the local potential [46]. Analogously the methane gas-exchange reaction provides carbon transfer to precursor particles according to adsorption, and diffusivity (Eq. 4) [46]:

$$CH_4 \rightleftharpoons [C]_{ad} + 3/2\, H_2 \qquad \text{(Eq. 4)}$$

Methane becomes unstable at roughly 800 K, but its decomposition temperature is lowered when dissociation occurs on a metal catalyst. For each set of carburizing or nitriding reactions, the gas flow was regulated using the LabVIEW-controlled gas flow regulators, where the reactive gas constituted 100 vol% of the chamber atmosphere following the initial purge. Previous work indicated that the application 100 vol% $CH_4$ or $NH_3$ aids conversion by maximizing the molecular availability of reactive species. More information about the thermodynamics and kinetics of gas-solid and laser-assisted reactions can be found elsewhere [22], [47]. With the reaction chamber reconstituted with reactive atmospheres, samples were selectively irradiated by rastering the beam over a 12 mm x 12 mm region of the powder bed according to a bi-directional raster pattern. Optical absorbance measurements of the metal and metal oxide precursor materials as they relate to photothermal energy transfer are presented in the supplementary section.

# 3 Results and Discussion

The ability to additively manufacture UHTCs using SLRS processing is contingent upon the complete and rapid reaction synthesis (on the orders seconds not hours like those of traditional UHTC carburization or nitridation reactions [32], [48], [49]) of the product ceramic while maintaining its material integrity. By irradiating the precursor materials using directed laser energy in the presence of 100 vol.% $CH_4$ or $NH_3$ gas, reaction-bonded carbide (TiC, ZrC, and HfC) and nitride (TiN, ZrN, and HfN) materials were obtained. Laser processing in Ar was used as a reference for the response of the material to inert versus reactive processing. The conversion of single-component precursors using initially fixed processing parameters was conducted (3 W, 100 mm/min unless otherwise noted) to examine each reaction and better understand the thermodynamic and kinetic factors that influence conversion, particle adhesion, volumetric changes, and microstructure. Optical micrographs showing the surface microstructure of each metal or metal oxide sample are presented in Fig. 5. Phase composition results from SLRS processing are presented for clarity of discussion according to their metal and accompanying metal oxide (Ti/$TiO_2$, Hf/$HfO_2$, and Zr/$ZrO_2$). Note that given the depth-dependent heating profile and precursor conversion, the phase fraction of non-oxide product in the uppermost region of the characterized layer may be greater than the XRD-averaged measurement. The results for the conversion of metal or metal oxide precursor systems are summarized in Tables 7 and 8

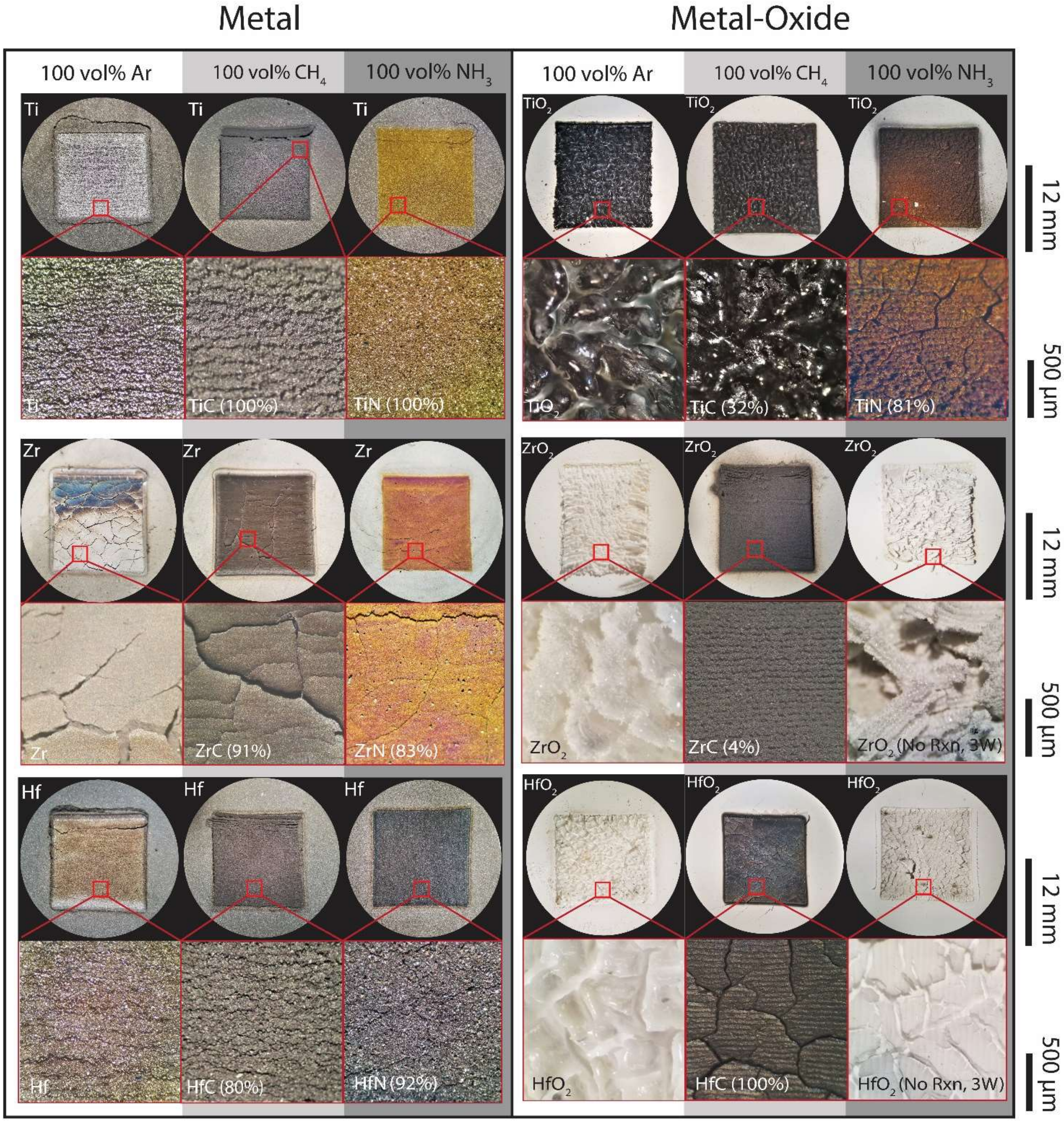

**Figure 4**. Photomicrographs of metal or metal oxide precursor powders irradiated by the 3.0 W laser at 100% power (100 mm/min scan speed) in 100 vol.% Ar (control), $CH_4$, $NH_3$. Percentages and target non-oxide products are depicted in the lower right corner of each image and indicate the total wt% of carbides or nitrides (and/or oxycarbides and oxynitrides) formed.

## 3.1 Selective Laser Reaction Synthesis of TiC and TiN

### *3.1.1 Conversion of Ti and $TiO_2$ Precursors to TiC by $CH_4$*

Carbothermal reduction and conversion of metal and metal oxide materials are common using traditional furnace heating, but a smaller body of literature exists on the synthesis of UHTC materials using laser-assisted methodologies in carbonaceous gaseous atmospheres. Several studies have indicated the direct laser synthesis of TiC using pulsed laser processing of bulk Ti in $CH_4$ [50], [51], *in-situ* TiC formation using selective laser melting of Ti and $B_4C$ [52], the synthesis of TiC thin films by pulsed laser deposition using Ti in $CH_4$ [53], and laser pyrolysis of $TiO_2$ in liquid titanium isopropoxide [54], yet direct laser-processing of precursor powders for AM has not been reported. By irradiating commercially available Ti and $TiO_2$ precursor materials using the diode laser, $TiC_y$ phases were produced.

Solid solutions of TiC-TiO have the same cubic crystalline structure as TiC so there are no additional reflections that reveal vacancies, oxygen or carbon in titanium interstitial sites (this is also the case for other NaCl-type UHTCs) [55]. Interstitial concentrations of carbon (nitrogen), or oxygen are related to both the physicochemical properties of the material and dictate the lattice parameter of interstitial carbide, oxycarbide, nitride and oxynitride species [55], [56]. Lattice parameter variations are attributed to the level of vacancies present in the metal and non-metal lattices and can be measured using Rietveld refinement and compared to established parameter-composition relationships to estimate the composition of non-stoichiometric carbide and oxycarbides produced by SLRS. The XRD spectra and carbide yields for conversion of Ti and $TiO_2$ are indicated Figure 5a, b, as well as near net-shape mixtures of metal/metal oxides.

Quantitative phase characterization indicated up to >99.9 wt% $TiC_{1.0}$ and 32.2% $TiC_{0.76}O_{0.25}$ were produced from the conversion of Ti and $TiO_2$, respectively [55]. No significant traces of carbon residue or deposition were detected suggesting the rate of reactivity was greater than the rate of $CH_4$ decomposition in the processing temperature regime. A limited comparison of the product yields from Ti and $TiO_2$ can be made. It might be expected that smaller particles with greater surface-area-to-volume ratios ($TiO_2$ for the results here) are typically associated with greater conversion yet the disparity in $TiC_y$ yield between the Ti and $TiO_2$ precursor materials is evident. Higher conversion of the Ti compared to the oxide precursor may be due to a variety of factors including the higher reaction temperature required for spontaneous conversion of $TiO_2$ as well as its extremely low optical absorption at 445 nm, which was the lowest of all precursor materials used in this study. Optical microscopy of $TiO_2$ processed in Ar indicated that physiochemical changes occurred independently of reactivity with $CH_4$ (or $NH_3$). The surface microstructure of irradiated $TiO_2$ was significantly consolidated and appeared to have reached temperatures near the melting point of $TiO_2$ (1843 °C). Most notably, black $TiO_2$ was indicated as a primary reaction product when processed in Ar and at higher laser energy densities. Black $TiO_2$ results from the formation of defects or partial reduction and has been noted to give the $TiO_2$ photoactivity and absorption in the visible [57]. This increase in absorptivity may result in a positive feedback loop where both defect formation and non-oxide synthesis can facilitate greater local temperatures, thereby further accelerating reactivity to non-oxide phases.

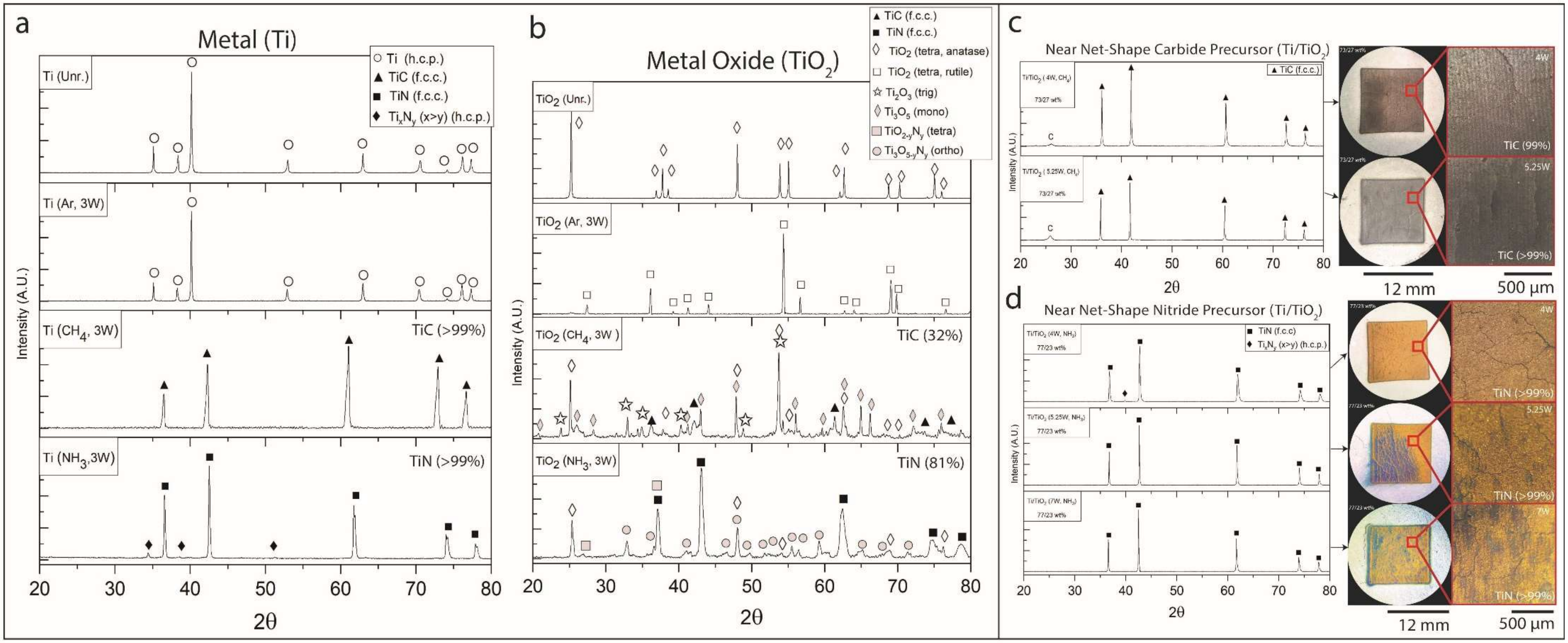

**Figure 5.** (a, b) XRD spectra of Ti and $TiO_2$ precursor materials that were processed using selective laser reaction sintering in >99.9 vol.% Ar, $CH_4$, and $NH_3$ using the 445 nm diode laser. (c, d) XRD spectra of near net-shape Ti/$TiO_2$ precursor materials that were processed using selective laser reaction sintering in >99.9 vol.% Ar, $CH_4$, and $NH_3$ using the 445 nm diode laser Photomicrographs of laser-processed samples are depicted adjacent to each XRD spectra for near net-shape compositions of metal/metal oxides that were used to reduce volume change induced stress.

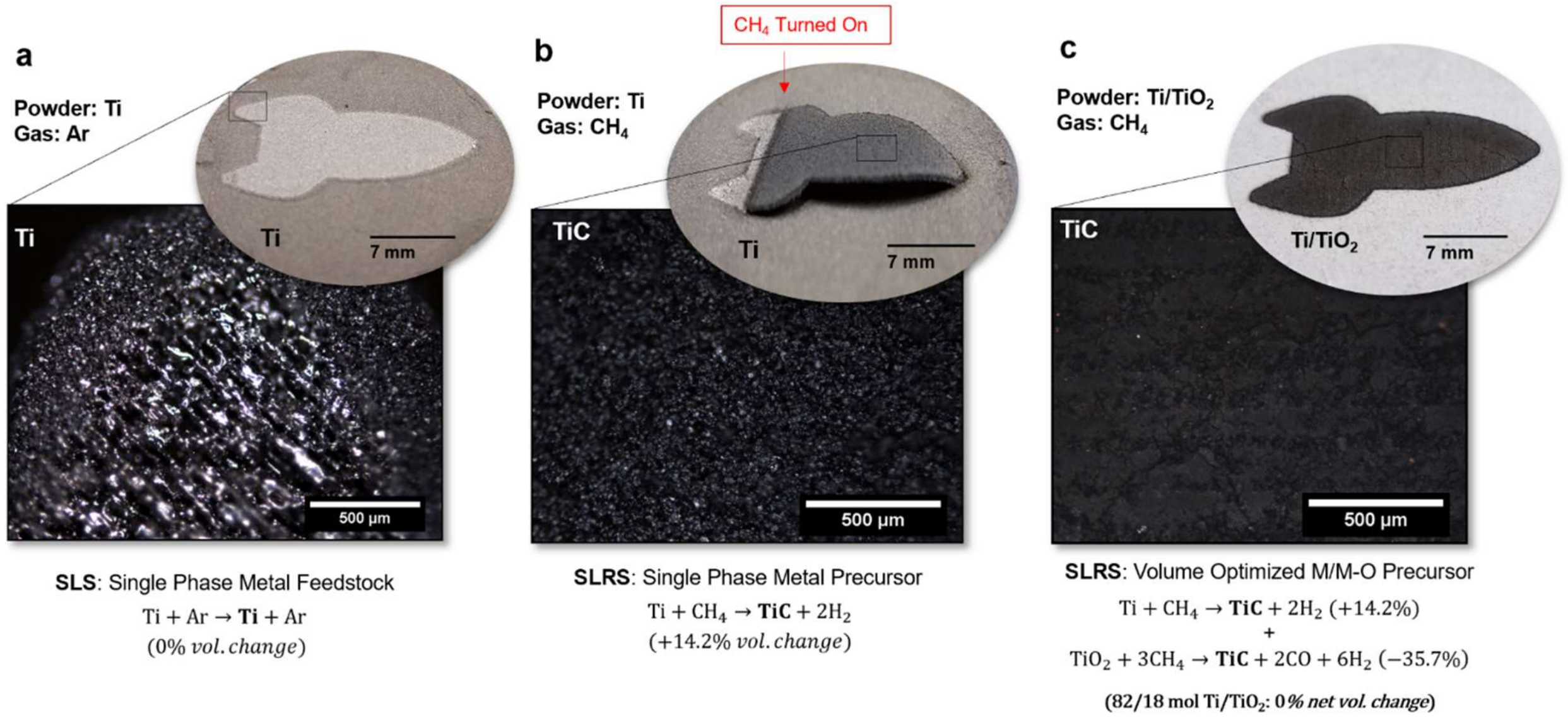

**Figure 6.** Comparison between different powder bed fusion style AM approaches. (a) SLS processing of Ti feedstock in Ar into a "rocket ship geometry" with the sintered Ti microstructure. (b) SLRS processing of the same Ti feedstock in $CH_4$ gas resulted in the *in-situ* conversion of Ti to TiC and volumetric expansion. The expansion resulted in residual stress and caused the rocket-shaped layer to bow off of the powder bed surface. (c) SLRS of near-net-shape 82/18 mol. Ti/$TiO_2$ precursor material, where volume changes or each precursor material compensate for one another to produce an even TiC layer via reaction bonding.

The diffraction data suggest that Ti was readily and directly converted to TiC, where volumetric expansion surpassed the degree of powder bed consolidation during inert SLS processing. The +14.2% expansion (Ti to $TiC_{1.0}$) of the single-phase metal precursor caused the SLRS layer to expand ultimately leading to the curvature depicted in Fig. 6. By contrast, conversion of $TiO_2$ to TiC was obtained through a more circuitous, multi-step reduction pathway that was associated with the generation of porosity in the converted structure. Fig. 6 indicates that concurrent with conversion by $CH_4$ or ($NH_3$), $TiO_2$ passes through a martensitic transformation from the tetragonal polymorphs of the anatase phase to the rutile phase. This transformation is noted to occur between 500 and 800 °C and gives some indication of the minimum local temperatures achieved during processing [58]. Evidently, during SLRS of TiC, anatase-$TiO_2$ (26.9 wt%), rather than rutile-$TiO_2$ was the residual oxide phase remaining in the irradiated layer. Others note that for the isothermal reduction of $TiO_2$ in $CH_4/H_2$ atmospheres, TiC formation occurs through the following sequence at a suggested optimal temperature of roughly 1300-1450 °C [59], [60]:

$$TiO_{2-z} \rightarrow Ti_5O_{9-z} \rightarrow Ti_4O_{7-z} \rightarrow Ti_3O_{5-z} \rightarrow Ti_2O_{3-z} \rightarrow (TiC_y + TiO_{1-z})_n \qquad \text{(Eq. 5)}$$

$$\tfrac{1}{2}\,Ti_2O_{3-z} + (\tfrac{1}{2} - 2x)CH_4 \rightarrow xTiC_{y\,(ss)} + (1-x)TiO_{z\,(ss)} + (\tfrac{1}{2} - x)CO - (1+4x)H_2 \qquad \text{(Eq. 6)}$$

The data indicate that rutile-$TiO_2$ was readily reduced by $CH_4$ to other sub-oxide phases (23.7 wt% $Ti_3O_5$, 17.1 wt% $Ti_2O_3$) during processing, where cubic $TiC_{0.76}O_{0.24}$ was formed from $Ti_3O_5$ or $Ti_2O_3$. The presence of these phases, as indicated by XRD, is broadly consistent with the prevailing furnace reduction mechanism despite laser-assisted conversion occurring locally on timescales <1 second (compared to hours). Burger noted that during reduction, CO reduction products simultaneously act as a reducing agent and are disproportionate at the surface wherein carbon is then incorporated into and diffuses into the crystal lattice [61]. Oxide particles are therefore precursors to oxycarbides where the oxygen rate loss and carbon incorporation are not equal. This leads to the production of sub-stoichiometric $TiC_xO_y$ [61]. These XRD measurements illustrate that there is a gradual transition of the oxide to oxycarbides crystal structure from a vacancy containing NaCl-type structure of TiO to fully occupied TiC as carbon content increases [55]. Similarly, the presence of oxygen in TiO (with stoichiometry varying from $TiO_{0.75}$ to $TiO_{1.3}$) shifts the largest intensity 2θ peak of TiC from 41.8° towards 43.2° as oxygen concentration increases [8], [55]. An analogous gradual mechanism for the carbidization of Ti occurs as lattice sites are filled and stoichiometry of the carbon sublattice approaches $TiC_{1.0}$. According to Zhang [59], by utilizing $CH_4$-rich gas compositions, and higher processing temperatures, the conditions for increased mass transfer of oxygen and carbon within solid solution could be optimized such that the TiC concentration approaches unity during the SLRS conversion of the Ti/$TiO_2$ precursor system. SLRS processing parameters (laser power, scan speed, atmosphere) must be tuned appropriately to fully reduce metal oxide species while simultaneously facilitating thorough conversion of the metal

### *3.1.2 Conversion of Ti and $TiO_2$ Precursors to TiN by $NH_3$*

Laser-assisted nitridation for surface modification of bulk and coating materials has been thoroughly examined for surface or tribological property modification of Ti [22], [62]. In particular, nitridation of Ti to TiN (by gas-phase $N_2$ [24], [26], [27], [63] or by liquid $N_2$ [64]) is one of the most widely examined systems for laser nitridation of bulk Ti materials. However, this process has not been extensively investigated for reactive powder bed fusion methodologies. Similar to other nitride materials the mechanical and physiochemical properties of the nitride strongly depend on the nitriding parameters and working atmospheres [65]. Several gases including $NH_3$, $H_2$-$N_2$, $N_2$, and Ar-$N_2$ have been used as nitriding atmospheres.

The results presented here for SLRS processing of Ti in >99.9 vol% $NH_3$ indicate that the formation of TiN was nearly complete using the power density of 8 W/mm$^2$. Two-phase compositions for nitrides were found: NaCl-type $TiN_{0.96}$, and hexagonal-$Ti_xN_y$ (x>y; ~$TiN_{<0.22}$). This result agrees with work by Khatko who observed that when the temperature in the irradiated zone is less the melting temperature of the metal, a combination of NaCl-type TiN and hexagonal $Ti_xN_y$ phases were formed during the high power (and temperature), pulsed laser processing of Ti in nitrogen saturated environments ($10^2$-$10^7$ W/mm$^2$ which is about 2-5 orders of magnitude greater intensity than used in this study) [64]. Peak temperatures achieved during irradiation Ti were not determined; Ti transforms from α (h.c.p.) to β (b.c.c.)-phases at approximately 800 °C, no but β-Ti was detected in the diffractogram (from 3-7 W laser power) despite partial melting of precursors when irradiated with the maximum laser power [66]. According to the Ti-N phase diagram, the presence of $TiN_{<0.22}$ (hexagonal close-packed) indicates that interlattice nitrogen in α-Ti increases rapidly during non-isothermal heating. The result is that NaCl-type $TiN_y$ may be formed directly from α-Ti-N, where the high-temperature β-Ti-N phase is not likely to be an intermediate [23], [67].

Conversion-induced volumetric expansion and porosity reduction during laser gas nitridation of Ti was observed by Wu and colleagues where layers of laser nitride TiN had a denser microstructure than the original material [27]. Figure 7 shows SEM micrographs of the Ti and $TiO_2$ samples laser processed in $NH_3$. A comparison between laser sintered Ti and the TiN product indicates that the microstructure of the nitride layers appeared denser than the as-deposited Ti material where small overlapping cracks are indicative of stress/pressure ridges. Even so, little reaction sintering (that might be applicable to AM) was indicated. The results indicate rapid conversion to ~94 wt% TiN occurred before significant diffusion was achieved. Under short reaction times, diffusion is limited and impacts particle bonding. Particles appeared nearly discretely converted and the small volume change upon conversion (~11.6 vol% expansion for stoichiometric yield) seemed to be accommodated by powder bed porosity. As a result, the converted powder layer had no appreciable mechanical integrity and could not be removed from the powder bed.

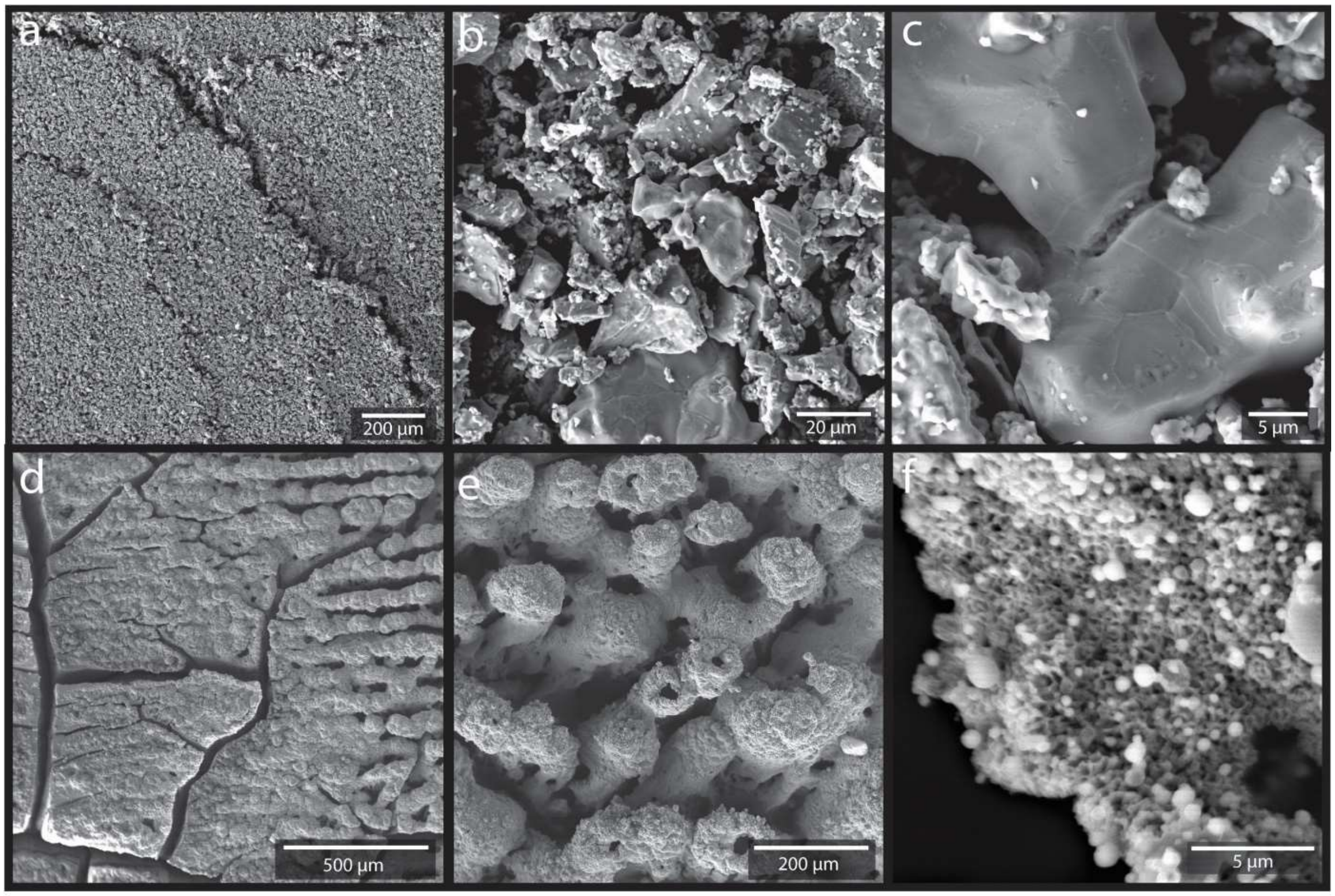

**Figure 7.** SEM micrographs of selective laser processed Ti (a-c) and $TiO_2$ (d-f) in >99.9 vol% $NH_3$ using 3W laser power.

Examination of $TiO_2$ processed in $NH_3$ indicates that, unlike SLRS of Ti, reduction of the metal oxide was largely incomplete under the same reaction conditions but samples had a characteristically different microstructure. The reaction of $TiO_2$ with $NH_3$, yielded up to 81.4 wt% total oxynitrides $TiN_yO_z$ phases, with the broadening of TiN peaks through the formation of $TiN_{0.35}O_{0.65}$ observed at 42.8° and 62.0° 2θ. Evidently, nitridation in $NH_3$ proceeded through similar reduction mechanisms as discussed in the synthesis of TiC from $TiO_2$. Distinct titanium oxynitride phases were evident in the XRD diffractogram in the form of $Ti_2O_{3-y}N_y$ and $Ti_3O_{5-y}N_y$. The presence of oxynitride has different thermodynamic stability compared to oxycarbide materials and indicates that conversion was not entirely achieved using the low power reaction parameters [68]. The stoichiometry of the NaCl- type product from the reduction of $TiO_2$ in $NH_3$ was estimated to be $TiN_{0.35}O_{0.65}$ using the 3 W laser power. Increased reaction temperatures would yield a more complete reduction to TiN. SEM micrographs in Figure 7d-e show the formation of sintered, off-gassing pores indicative of external diffusion of gaseous by-products such as $H_2O$. The particles in the SLRS processed $TiO_2$ layer appeared significantly consolidated and well-bonded. Additionally, long-range mud-cracking was observed due to a combination of sintering-induced densification and conversion-related volume reduction.

## 3.2 Selective Laser Reaction Synthesis of ZrC and ZrN

### *3.2.1 Conversion of Zr and $ZrO_2$ Precursors to ZrC by $CH_4$*

According to Ushakov [9], interest in ZrC for extreme high-temperature environment applications is increasing rapidly. There are more publications on ZrC per year than for ZrN, HfC, and HfN combined, largely due to the need for barrier coatings and high-temperature materials for nuclear reactors [9]. Many synthetic approaches established for TiN and TiC can be adapted for Zr and Hf carbides and nitrides [9]. There are numerous processing strategies to form ZrC, of which carbothermal reduction using solid carbon is most common for powder synthesis (carbothermal reduction can be achieved using traditional heating [69], laser pyrolysis [70], or reactive spark plasma sintering [8], [48], [69]). In addition to the preparation of ZrC through oxide reduction, ZrC and ZrN can be obtained through the direct, self-propagating reaction of metal or metal hydride powders with graphite [8]. Yet, an expeditious ZrC AM-compatible synthesis approach using chemical vapor deposition reactions induced by laser irradiation has not been widely studied.

X-ray phase composition results indicate that SLRS of Zr in $CH_4$ (using 3 W laser powder) yielded approximately 90.8 wt% $ZrC_x$ with unreacted Zr (2.9 wt%) and $ZrO_2$ (6.6 wt%) due to oxidation remaining as impurity phases. Because oxidation of Zr is more thermodynamically favorable than carbonization, lattice parameter results indicate the presence of competitive oxidation during SLRS conversion. Ursu noted that during laser processing of Zr, oxides appear when oxygen contamination exceeds $2\times10^{-4}$ and should be kept below this level for powder bed fusion techniques [26]. As a result, the $ZrC_y$/ $ZrC_yO_{1-y}$ stoichiometry may be $ZrC_{0.89}O_{0.11}$ rather than the $ZrC_{0.75}$ since both compositions have the same lattice parameter (C increases the lattice parameter, O decreases it). The combined volume change associated with this conversion (+10.65% for Zr to $ZrC_{0.89}O_{0.11}$ computed by lattice parameter change and 54.1% for oxidation of Zr to $ZrO_2$) caused the layer to bow off the powder bed surface.

By contrast to the conversion of Zr, the reduction of $ZrO_2$ necessitates higher processing temperatures for spontaneous conversion (~685 °C). During SLRS processing, this thermodynamic threshold was achieved through photothermal irradiation despite the low levels of optical absorption for $ZrO_2$ at 445 nm ($3.3\times10^{-3}$ a.u. ). Overall laser processing of $ZrO_2$ in $CH_4$ produced low carbide yields of approximately 3.5 wt% $ZrC_yO_{1-y}$ using the 3 W processing conditions. During SLRS synthesis, the majority of the unreacted oxide phase was determined to be monoclinic-$ZrO_2$ (95.6 wt%) with approximately 0.9 wt% remaining with a tetragonal crystal structure. Undoped monoclinic-$ZrO_2$ undergoes a temperature-induced martensitic transformation from monoclinic to tetragonal polymorphs at approximately 1170 °C and is stable up to 2370 °C at which point cubic zirconia is formed [48], [70]. XRD results suggest that both the spontaneous reaction temperature for ZrC formation (685 °C) and the martensitic transformation temperature were reached during SLRS where peak reaction temperatures occurred (somewhere between 1170 °C and 2370 °C). This relatively low-temperature processing may have produced inefficient reaction kinetics for rapid conversion even if thermodynamic reaction thresholds were exceeded.

Laser power was increased to 4 W, (~32W/$mm^2$*)*, and 5.25 W (~58W/$mm^2$) to reach the high temperatures needed for conversion. Rietveld results using 4 W laser power yielded 56.1 wt% $ZrC_{0.93}O_{0.07}$, monoclinic-$ZrO_2$ (43.6 wt%), and trace tetragonal-$ZrO_2$ (<0.3 wt%). Carbothermal reduction mechanisms for metal oxides with low volatility oxide intermediates ($ZrO_2$ and $TiO_2$) occur through the formation of oxycarbides intermediates (e.g. $ZrO_xC_y$, $TiO_xC_y$). Using either set of processing conditions, no distinct cubic-ZrO phases akin to $TiO_{0.75}$-$TiO_{1.3}$ were indicated by the XRD. This result corroborates previous studies where the cubic rocksalt structure of ZrO was not unambiguously confirmed [8], [61]. Still deviations in stoichiometry from $ZrC_y$ due to the substitution of carbon for oxygen mark a noticeable decrease in lattice parameters [61], [71]. XRD phase characterization indicates cubic-$ZrC_{0.88}O_{0.12}$, $ZrC_{0.91}O_{0.09}$ and $ZrC_{0.93}O_{0.07}$ were produced from 3 W, 4 W, 5.25 W processing conditions respectively with increasing oxycarbide yield. Residual oxygen substitution is indicative of incomplete reduction and is noted to influence the sintering and densification mechanisms [32]. The differences related to peak processing temperature are not known, but the significant increase of $ZrC_y$ -$ZrC_yO_z$ yield across the relatively narrow range of processing conditions suggests that the production of carbide species accelerates rapidly with increasing temperatures [48].

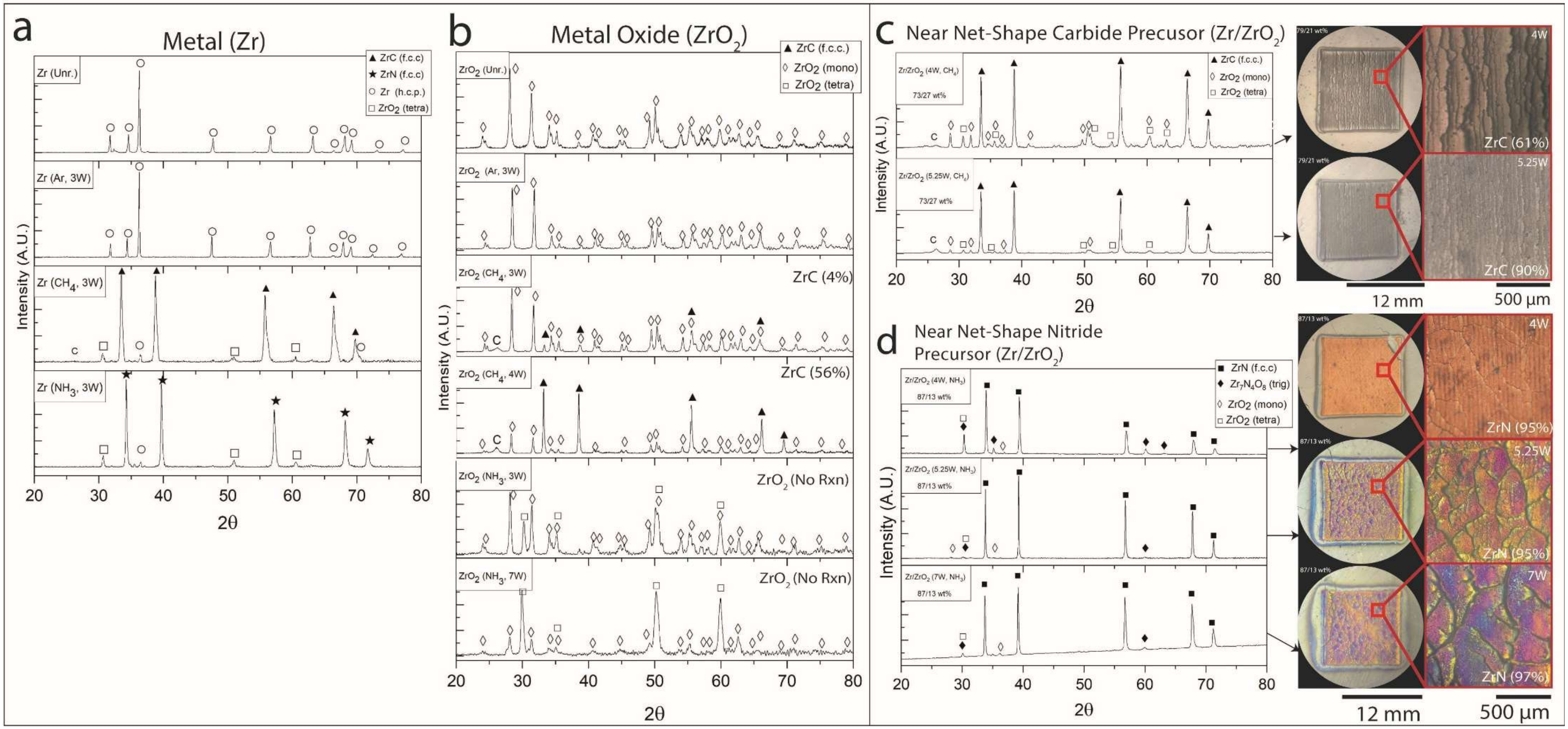

**Figure 8.** (a, b) XRD spectra of Zr and $ZrO_2$ precursor materials that were processed using selective laser reaction sintering in >99.9 vol.% Ar, $CH_4$, and $NH_3$ using the 445 nm diode laser. (c, d) XRD spectra of near net-shape $Zr/ZrO_2$ precursor materials that were processed using selective laser reaction sintering in >99.9 vol.% Ar, $CH_4$, and $NH_3$ using the 445 nm diode laser Photomicrographs of laser-processed samples are depicted adjacent to each XRD spectra for near net-shape compositions of metal/metal oxides that were used to reduce volume change induced stress.

### *3.2.2 Conversion of Zr and $ZrO_2$ Precursors to ZrN by $NH_3$*

The Zr-N system is characterized by only one nitride phase up to the stoichiometric composition ZrN, and one super-stoichiometric $Zr_3N_4$ phase [7]. Zirconium nitride can accommodate non-metal vacancies thereby producing non-stoichiometric $ZrN_{1-x}$ or $ZrN_yO_{1-y}$ when heated in the presence of nitrogen gas [49]. Similarly, ZrN can be obtained through the direct, strongly exothermic, self-propagating reaction of metal or metal hydride powders with $N_2$ or $NH_3$ or by reduction synthesis of $ZrO_2$ in $N_2$ or $NH_3$ [8], [45]. A series of studies by Ursu and colleagues demonstrated that laser surface modification of Zr in $N_2$ produces ZrN; however the study was limited to coating formation on bulk materials [23]–[26], [72]. XRD analysis of Zr SLRS processed in $NH_3$ (shown in Figure 8.) indicates 82.6 wt% $ZrN_y$ and tetragonal-$ZrO_2$ (11.9 wt% due to oxidation) were produced with unreacted Zr (6.0 wt%) remaining. Cubic-$ZrN_y$ was the only nitride product observed. This contrasts with previously reported results where a combination of cubic-$ZrN_x$ and orthorhombic-$Zr_2N$ were formed [64]. According to the literature, laser gas nitridation in $N_2$ can induce significant competition between nitridation and oxidation if oxidizing impurities are present in the gaseous atmosphere [23]. Although the reaction chamber in this investigation was purged with argon for 2 hrs before laser processing, the high surface-area-to-volume ratio of Zr particles made precursors extremely susceptible to the formation of $ZrO_2$ [73]. The gold-red color of the $ZrN_{0.25}O_{0.25}$ product is indicative of its stoichiometry (Figure 8d). Sintered layers become yellower and redder with increasing nitrogen content; meanwhile with increasing oxygen content the color shifts from lighter to a darker, less vibrant yellow [74], [75]. These changes are characterized by the behavior of free electrons, where color variation of ZrN is observable by eye if the atomic ratio is changed by 0.03 [74], [75]. In short, quantitative (and qualitative) phase characterization of SLRS processed Zr → $ZrN_y$ suggests that because Zr thermodynamically prefers to react with oxygen than nitrogen at low temperatures [75], oxidation competes with nitridation during SLRS processing even if trace impurities are present. Also, results show that $ZrN_y$ synthesis proceeds rapidly during exothermic laser synthesis.

In contrast to the laser-assisted conversion of Zr in $NH_3$, the reduction of $ZrO_2$ did not produce ZrN phases under the prevailing reaction conditions. Low optical absorption in the precursor at 445 nm ($3.3\times10^{-3}$ a.u.), coupled with high thermodynamic thresholds for reactivity (2008 °C, the highest in this study) inhibited rapid conversion. According to Kathuria [44], during laser surface nitridation of yttria-stabilized tetragonal-$ZrO_2$ (t-$ZrO_2$) in $N_2$, crystallographic modifications occur during nitridation where t-$ZrO_2$→c-$ZrO_2$→ZrN with increasing nitrogen concentration. Even though tetragonal-$ZrO_2$ was present in the diffractogram of $ZrO_2$ (forms at 1170 °C), no indication of cubic-$ZrO_2$ (forms at 2370 °C) or ZrN was detected when the maximum laser power (7 W) was used. The synthesis of ZrN from $ZrO_2$ via laser nitridation has been demonstrated by others but likely required significantly higher laser intensities than were used in this study [76]. The results indicate that higher temperatures are required for SRLS of precursor mixtures containing $ZrO_2$. An increase in precursor optical absorption and/or laser power density could facilitate increased temperatures and reactivity.

## 3.3 Selective Laser Reaction Synthesis of HfC and HfN

### *3.3.1 Conversion of Hf and $HfO_2$ Precursors to HfC by $CH_4$*

The prevailing practices for HfC synthesis involve the carbothermal reduction of nanoscale $HfO_2$ particles with graphite or carbon black over multi-hour-long syntheses [28]. No investigations on rapid laser-assisted processing of HfC have been reported (AM or otherwise). Selective laser processing of Hf in $CH_4$ (3 W) produced 79.5 wt% cubic-$HfC_{0.84}$ with 20.5 wt% Hf remaining as unreacted precursor material. No oxide peaks were found in the diffraction spectra suggesting that oxygen substitution within the Hf lattice was low. An amorphous carbon peak at approximately 26° 2θ in the diffractogram indicated that trace carbon impurities were present on the surface of the HfC-containing SRLS layer. Carbon residue appeared without full conversion. This may suggest the rate of local carbon absorption and/or gas-phase decomposition and nucleation exceeded the rate of interparticle diffusion. The formation of a $HfC_y$ passivation layer on incompletely reacted particles might have prevented the reaction from going to completion [21], [77]–[81]. According to thermodynamic calculations, the conversion of Hf by $CH_4$ is exothermic and occurs spontaneously at 319 °C which is the lowest of carbide synthesis temperatures for the metallic precursors investigated in this study. Hf precursor particle sizes were roughly an order of magnitude larger than Zr. High optical energy absorption (0.52 a.u. at 445 nm) and exothermic reaction conditions facilitated rapid $HfC_y$ formation under the prevailing reaction conditions and scan parameters (100 mm/min). These results using larger particles are promising for application to selective laser reaction sintering AM where the typical average particle size of SLM and SLS feedstock materials is 10-50 μm [82]. Analogous to the conversion of Ti →TiC and Zr→ ZrC, the carbidization of Hf resulted in net volume expansion that caused the reaction sintered layer to bow off the powder bed and produced stress/pressure ridges shown in Figure 9a, b.

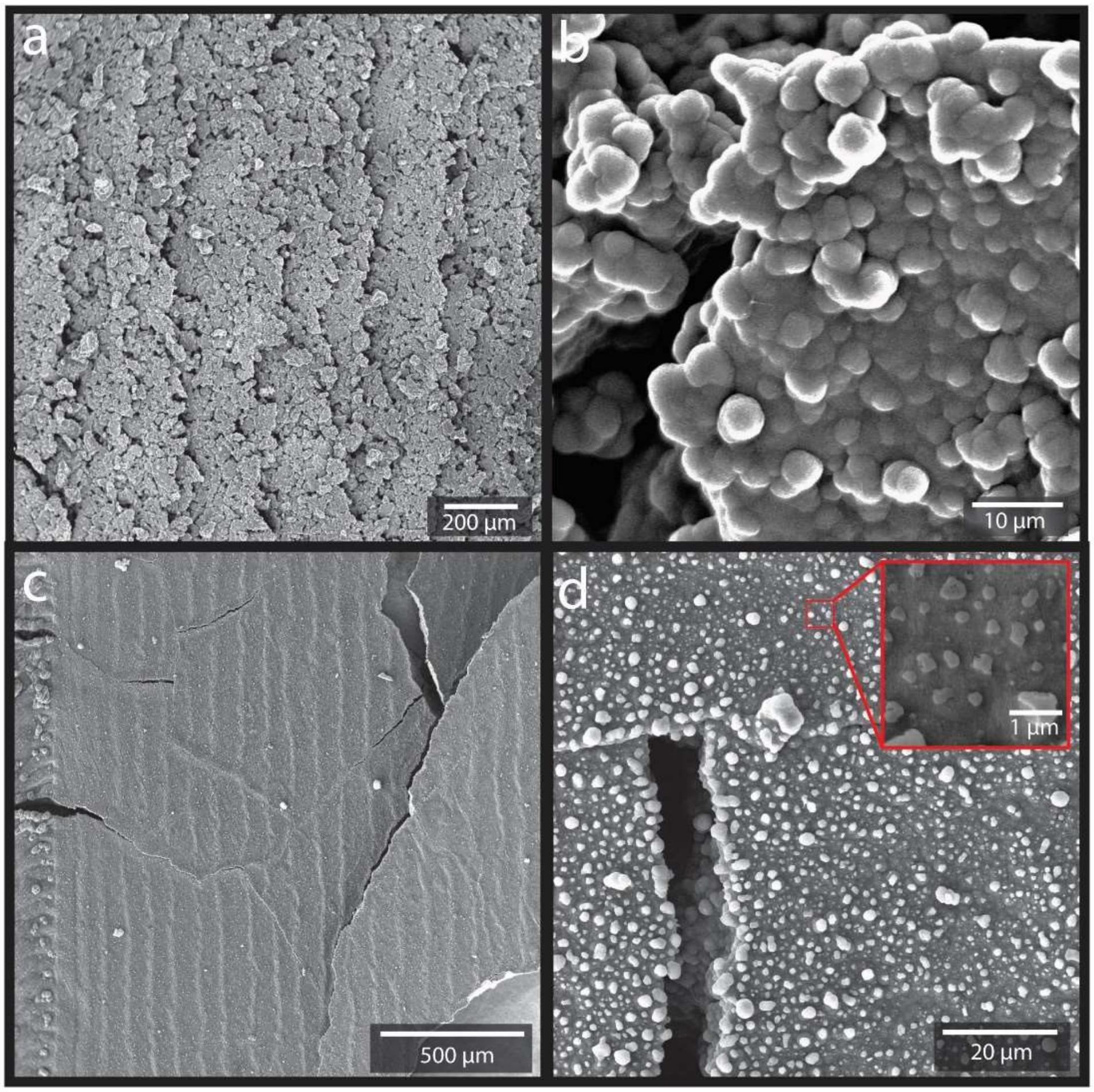

**Figure 9.** SEM micrographs of selective laser processed Hf (a,b) and $HfO_2$ (c,d) in >99.9 vol% $CH_4$ using 4W laser power.

$HfO_2$ and $ZrO_2$ are known as "oxide twins" due to their similar crystal structures, and martensitic polymorphs (monoclinic, tetragonal, cubic). Like $ZrO_2$, $HfO_2$ undergoes temperature-dependent martensitic transformations [83]:

$$\text{m-HfO}_2 \xrightarrow{1720°C} \text{t-HfO}_2 \xrightarrow{2600°C} \text{c-HfO}_2 \qquad \text{(Eq. 7)}$$

The prevailing theory suggests they are reduced through similar reaction mechanisms [9], [83]. A comparison of the diffractograms for unreacted precursor materials and materials laser-processed in Ar illustrates that the martensitic transformation temperature of 1720 °C was reached. Temperatures associated with the monoclinic to tetragonal polymorphic transformation $HfO_2$ greatly exceed reduction temperatures and would create favorable reaction conditions for HfC synthesis. Rietveld refinement indicated that nearly >99.9 wt% cubic-$HfC_{1.0}$ or $HfC_{0.97}O_{0.02}$ was produced via SLRS processing of $HfO_2$. Studies on the carbothermal reduction $HfO_2$ (using solid carbon) indicate that during heating up to 1400 °C removal of oxygen leads to C diffusion into the hafnia lattice giving monoclinic hafnium oxycarbides $HfC_yO_{2(1-y)}$. Further heating to 1600 °C causes transformation to cubic-oxycarbides ($HfC_yO_{(1-y)}$) [28]. High temperatures allow for faster atomic diffusion permitting reactions to approach completion over relatively short times. Figure 10 suggests that SLRS-induced reduction of $HfO_2$ only resulted in cubic, $HfC_xO_{(1-x)}$ or $HfC_x$ product materials, without the formation of monoclinic hafnium oxycarbides $HfC_xO_{2(1-x)}$. The conversion that occurred despite low optical absorbance ($2.6\times10^{-3}$ a.u.) could be the result of nonlinear optical absorption or gas-phase deposition of solid C onto $HfO_2$ particles that could absorb the laser energy and increase local temperatures. Microstructural characterization shown in Figure 9c,d indicates that the small $HfO_2$ particles were significantly consolidated and bonded upon conversion. A combination of reaction sintering-related consolidation, particle coalescence (to reduce surface free energy), and net volume reduction likely contributed to the formation of large cracks and voids. High-magnification SEM images show the formation of nucleated islands of material on the surface of the >99.9 wt% converted $HfC_{1.0}$ material. The exact composition of these features was not determined but their presence indicates significant surface diffusion on SLRS-UHTC layer concurrent with and following formation. Overall, high surface-area-to-volume ratios of nanoscale precursor particles and low thresholds for reactivity were sufficient enough to overcome low optical absorption characteristics facilitating high UHTC yield. The results indicate that nearly transparent materials can be rapidly converted to UHTCs using appropriate SLRS processing conditions.

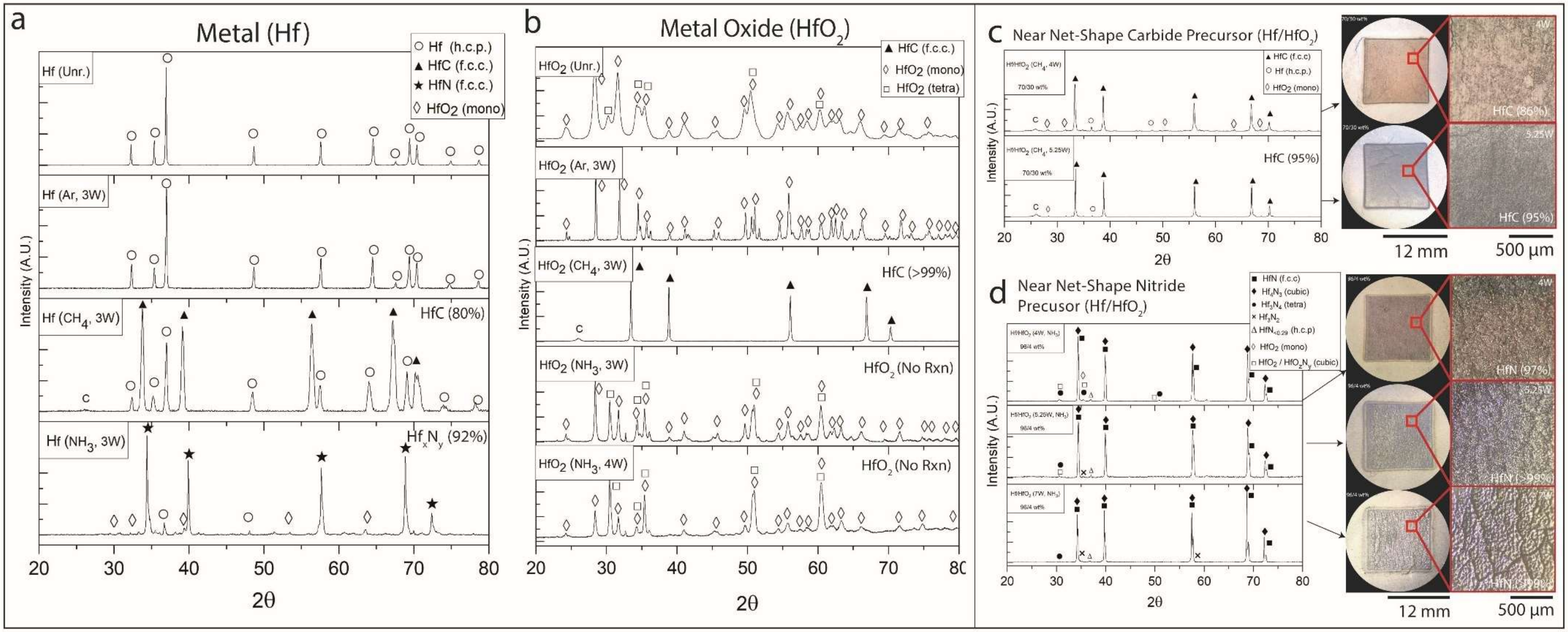

**Figure 10.** (a, b) XRD spectra of Hf and $HfO_2$ precursor materials that were processed using selective laser reaction sintering in >99.9 vol.% Ar, $CH_4$, and $NH_3$ using the 445 nm diode laser. (c, d) XRD spectra of near net-shape Hf/$HfO_2$ precursor materials that were processed using selective laser reaction sintering in >99.9 vol.% Ar, $CH_4$, and $NH_3$ using the 445 nm diode laser Photomicrographs of laser-processed samples are depicted adjacent to each XRD spectra for near net-shape compositions of metal/metal oxides that were used to reduce volume change induced stres

### *3.3.2 Conversion of Hf and $HfO_2$ Precursors to HfN by $NH_3$*

Unlike laser-assisted synthesis of $HfC_y$, synthesis of $HfN_y$ has been performed using surface treatments of bulk Hf or $HfO_2$ in $N_2$ (liquid) or $NH_3$ environments using high-power lasers ($10^2$-$10^7$ W/mm$^2$ ) [25], [64], [72], but synthesis techniques for reactive additive manufacturing applications using powdered precursor chemistries have not been reported. Quantitative phase characterization results indicated that the selective laser reaction sintering of Hf powder in $NH_3$ using the low power processing setting facilitated significant conversion to $Hf_xN_y$ (total 97.4%) with several nitride crystal structures formed ($Hf_3N_4$, $HfN_{\sim1.0}$, $Hf_4N_3$ $Hf_3N_2$, $HfN_{0.25}$). A small fraction of Hf precursor remained with no apparent oxidation being indicated by the XRD diffractogram. Ursu noted that the upper limit of oxygen concentration for laser nitridation of Hf is $5 \times 10^{-6}$ otherwise oxides phases form (compare to $2 \times 10^{-3}$ for Ti, $2 \times 10^{-4}$ for Zr). The use of large Hf particles, with lower surface-area-to-volume ratios, may contribute to the lack of oxidation formed. This suggests that particles in the 10-40 μm size range may be ideal for current additive manufacturing processing given their greater resistance to oxidation, relative flowability during PBF deposition, and sufficiently rapid conversion.

Broadly, the phase composition results in Figure 10 agree with the findings presented in previous studies on nitridation of bulk group IV transition metals. Results in this study as well as other reports on laser-assisted processing indicate that nitrogen content in reaction products decreases as a function of depth into the laser-processed zone. Moreover, the formation of the superstoichiometric orthorhombic-$Hf_3N_4$ phase at the surface of the precursor bed indicates that interstitial sites in the Hf lattice structure were filled supporting the result that near stoichiometric $HfN_{1.0}$ was obtained. Other $HfN_y$ phases including subnitrides $Hf_4N_3$, $Hf_3N_2$, and $HfN_{<0.29}$ were also formed. Rudy first reported the synthesis of $Hf_4N_3$ and $Hf_3N_2$ which were prepared from the reaction Hf metal with HfN and reported them to be stable below 1970 and 2300 °C, respectively [94]. During SLRS, the formation of nitrogen-rich phases at the surface of the powder bed may facilitate the conversion of adjacent Hf particles even if not directly irradiated so long as reaction temperatures are sufficiently high to enhance intraparticle and interparticle diffusion.

Results indicate that conversion conditions for the production of nitride phases from $HfO_2$ were not sufficient to produce a high yield of $HfN_y$. The minimum reaction temperature for the synthesis of HfN for $HfO_2$ was estimated to be ~1235 °C. Even though t-$HfO_2$ (forms at 1720 °C) was indicated by XRD , reaction kinetics and prevailing SLRS processing conditions using 3-4 W laser powers were not sufficient to induce desirable levels of conversion. Increasing laser power/energy densities could facilitate increased temperatures and reactivity.

**Table 6. X-Ray Composition Analysis Using Rietveld Refinement Modeling: Conversion of Metal Or Metal Oxide Precursors in $CH_4$ Using Laser Heating.**

| Target Product (Stoichiometric Lattice Parameter) | Solid Phase Precursor | Target Reaction | Stoichiometric Reaction with $CH_4$ | Laser Power (W) | Total Carbide and/or Oxycarbide (wt%) | NaCl-Type $M_xC_y$ or $M_xC_yO_z$, x≈ y+z (wt%) | $M_xC_y$ or $M_xC_yO_z$, Lattice Parameter (Å) | Calc. NaCl-Type Product Stoichiometry | Unr. M or C in α-M (wt%) | Unr. M-O (or oxidation) (wt%) | Carbon Residue |
|---|---|---|---|---|---|---|---|---|---|---|---|
| **TiC** | Ti | 1 | $Ti + CH_4 \rightarrow TiC + 2H_2$ | 3 | ~99.9 % | $TiC_y$: >99.9 % | 4.330 | ~$TiC_{1.0}$ * [55] | 0% | - | No |
| **4.327 Å** | $TiO_2$ | 2 | $TiO_2 + 3CH_4 \rightarrow TiC + 2CO + 6H_2$ | 3 | 32.2% | $TiC_yO_z$: 32.2% | 4.294 | $TiC_{0.76}O_{0.24}$ [34], [59], [84] | - | 67.8% | Yes |
| **ZrC** | Zr | 3 | $Zr + CH_4 \rightarrow ZrC + H_2$ | 3 | 90.8% | $ZrC_y$: 90.8% | 4.683 | $ZrC_{0.75}$* [71] (or $ZrC_{0.88}O_{0.12}$[71]) | 2.9% | 6.6% | Trace |
| **4.704 Å** | $ZrO_2$ | 4 | $ZrO_2 + 3CH_4 \rightarrow ZrC + 2CO + 6H_2$ | 3 | 3.5% | $ZrC_yO_z$ :3.5% | 4.688 | $ZrC_{0.91}O_{0.09}$ [85] | - | 96.5% | Trace |
| | $ZrO_2$ | 4 | $ZrO_2 + 3CH_4 \rightarrow ZrC + 2CO + 6H_2$ | 4† | 56.1% | $ZrC_yO_z$: 56.1% | 4.689 | $ZrC_{0.93}O_{0.07}$ [71], [85] | | 43.6% | Yes |
| **HfC** | Hf | 5 | $Hf + CH_4 \rightarrow HfC + H_2$ | 3 | 79.5% | $HfC_y$: 79.5% | 4.631 | $HfC_{0.85}$ * [86] | 20.5% | - | Yes |
| **4.639 Å** | $HfO_2$ | 6 | $HfO_2 + 3CH_4 \rightarrow HfC + 2CO + 6H_2$ | 3 | >99.9 % | $HfC_y$: >99.9 % | 4.639 | ~$HfC_{1.0}$* [86],[87] (or $HfC0_{.97}O_{0.03}$ [32]) | - | - | Yes |

*Carbide may contain trace oxygen contaminants and vacancies but results indicated concentrations were low. Citations for the relationship between lattice parameter and stoichiometry are noted for each NaCl type product phase [32], [34], [55], [56], [59], [71], [84]–[89]

## Table 7. X-Ray Composition Analysis Using Rietveld Refinement Modeling: Conversion of Metal Or Metal Oxide Precursors in $NH_3$ Using Laser Heating.

| Target Product (Stoichiometric Lattice Parameter) | Solid Phase Precursor | Target Reaction | Stoichiometric Reaction with $NH_3$ | Laser Power (W) | Total Nitride and/or Oxynitride (wt%) | $M_xN_y$, x>y (wt%) | NaCl-Type $M_xN_y$ or $M_xN_yO_z$, x≈ y+z (wt%) | $M_aN_b$, a>b (wt%) | $M_xO_yN_z$ (wt%) | $M_xN_y$ or $M_xN_yO_z$, Lattice Parameter (Å) | Calc. NaCl-Type Product Stoichiometry | Unr. M or N in α-M (wt%) | Unr. M-O (or oxidation) (wt%) |
|---|---|---|---|---|---|---|---|---|---|---|---|---|---|
| **TiN** | Ti | 9 | $2Ti + 2NH_3 \rightarrow 2TiN + 3H_2$ | 3 | >99.9 % | - | $TiN_y$: 93.9% | $TiN_{<0.22}$ 6.1% (N in α-Ti ) | - | 4.239 | $TiN_{0.96}$ [67] | 0% | - |
| 4.241 Å | $TiO_2$ | 10 | $6TiO_2 + 4NH_3 \rightarrow 6TiN + 12H_2O + N_2$ | 3 | 81.4% | - | $TiN_yO_z$: 59.7% | - | $Ti_3O_{5-y}N_y$; $Ti_2O_{3-y}N_y$ 24.4% (est.) | 4.223 | $TiN_{0.35}O_{0.65}$ [90] | - | 67.8% |
| **ZrN** | Zr | 11 | $2Zr + 2NH_3 \rightarrow 2ZrN + 3H_2$ | 3 | 82.6% | - | ZrN: 82.6% | - | - | 4.574 | $ZrN_{0.9}$ to $ZrN_{0.75}O_{0.25}$ [91] | 5.6% | 11.8% |
| 4.575 Å | $ZrO_2$ | 12 | $6ZrO_2 + 4NH_3 \rightarrow 6ZrN + 12H_2O + N_2$ | 3 | 0% | - | 0% | - | - | - | - | - | >99.9 % |
| | $ZrO_2$ | 12 | $6ZrO_2 + 4NH_3 \rightarrow 6ZrN + 12H_2O + N_2$ | 4 | 0% | - | 0% | - | - | - | - | - | >99.9 % |
| | $ZrO_2$ | 12 | $6ZrO_2 + 4NH_3 \rightarrow 6ZrN + 12H_2O + N_2$ | 5.25 | 0% | - | 0% | - | - | - | - | - | >99.9 % |
| | $ZrO_2$ | 12 | $6ZrO_2 + 4NH_3 \rightarrow 6ZrN + 12H_2O + N_2$ | 7† | 0% | - | 0% | - | - | - | - | - | >99.9 % |
| **HfN** | Hf | 13 | $2Hf + 2NH_3 \rightarrow 2HfN + 3H_2$ | 3 | 97.4% | $Hf_3N_4$ 11.0% | HfN: 11.4%, | $Hf_4N_3$: 60.2% $Hf_3N_2$: 8.9% $HfN_{<0.29}$: 5.9% | - | 4.523 | ~$HfN_{1.0}$ to $Hf_{1.1}$ [92] | 2.6% | - |
| 4.518 Å | $HfO_2$ | 14 | $6HfO_2 + 4NH_3 \rightarrow 6HfN + 12H_2O + N_2$ | 3 | 0% | | 0% | - | - | - | - | - | >99.9 % |
| | $HfO_2$ | 14 | $6HfO_2 + 4NH_3 \rightarrow 6HfN + 12H_2O + N_2$ | 4 | 0% | | 0% | - | - | - | - | - | >99.9 % |
| | $HfO_2$ | 14 | $6HfO_2 + 4NH_3 \rightarrow 6HfN + 12H_2O + N_2$ | 5.25† | 0% | | 0% | - | - | - | - | - | >99.9 % |
| | $HfO_2$ | 14 | $6HfO_2 + 4NH_3 \rightarrow 6HfN + 12H_2O + N_2$ | 7† | 0% | | 0% | - | - | - | - | - | >99.9 % |

†denotes that significant vaporization caused deposition of material on the quartz tube, reducing effective power. *Nitride may contain oxynitride species $M_aC_bO_x$ and or oxygen lattice substitutions. [55], [67]

## 3.4 Conversion of Near-Net Shape Metal/Metal-Oxide Precursor Materials

With a basic understanding of SLRS reactivity for each precursor material, laser energy density was guided by the phase and stoichiometric results in Table 6 and Table 7. Increased power density was used to increase local temperatures and facilitate reaction synthesis towards stoichiometric UHTC formation. For each of the materials systems (Ti/$TiO_2$, Zr/$ZrO_2$, Hf/$HfO_2$, Ta/$Ta_2O_5$), near-net-shape compositions were prepared according to their stoichiometric lattice parameters and bulk densities. Conversion results for the near-net-shape precursor mixtures are shown in Table 8 and Table 9. The phase composition results from the conversion of composite precursor systems are broadly consistent with the results from single component precursor materials where increased laser power density was associated with greater non-oxide product yield. Optical micrographs and XRD spectra for near net-shape precursor compositions are presented in Figs. 7, 9, and 11. Key differences for SLRS of composite precursors have significance for selective laser reaction sintering-AM and are described as follows:

1. Conversion of oxide species by local heating of the metal phase. The presence of highly absorbing metal particles increases materials absorptivity (compared to the oxide) according to a weighted rule of mixtures using the volume fraction of each component [93]. Results indicate that the absorptivity of the metal/metal oxide precursor mixture was sufficient to facilitate the conversion of both precursor species even if reaction conditions for processing the oxide alone were not favorable. For example, conversion of $HfO_2$ (4 W, medium power) in $NH_3$ did not yield efficient conversion to HfN. High reaction thresholds (~1235 °C), low optical absorption ($2.6\times10^{-3}$) and slow kinetics prevented conversion (0 wt% $Hf_xN_y$). By contrast processing of the 96/4 wt% Hf/$HfO_2$ precursor mixture indicated that the small fraction of $HfO_2$ was converted to $Hf_xN_y$, leaving only $HfO_2$ as the unreacted precursor phase. During photothermal heating of beds containing metal and metal oxide powders, light can penetrate the powder compact readily through the metal oxide phase and be absorbed by metal particles [21], [93], [94]. Photothermal heating of metal particles then facilitates local heating of the oxide particles percolated throughout the mixture via particle-particle heat conduction so that reaction thresholds can be reached.
2. Reduction of oxide particles by adjacent metal particles independent of reactant gas. Egziabher *et al.* noted that $TiO_2$ is reduced by Ti to form metastable $TiO_y$ when Ti/$TiO_2$ mixtures were processed at 900 °C in inert conditions [95]. Partial non-stoichiometric reduction to metastable species might facilitate more rapid conversion of transition metal oxides to non-oxides. This partial reduction by the metal may be accompanied by the vacancy formation (indicative of black $TiO_2$ formation and increased absorptivity) that results from processing using high laser energy densities. These combined effects might be responsible for the higher levels of conversion (for oxides) that occurred for near-net-shape precursor mixtures compared to the weighted average product produced by single-component materials. This interaction between metal and metal oxide constituents does impact the product microstructure. Conversion of the metal precursor is associated with volume expansion that in the absence of oxygen and with high laser energies can increase product density [27]. As the local oxide content increases, reacted structures form a spongy, spalled exterior despite an increase in net occupancy of material [26]. Evidence of this effect is shown in Fig. 12 and may accelerate conversion by decreasing the difficulty of reactant penetration but also might result in unintended porosity in product structures under certain processing regimes.

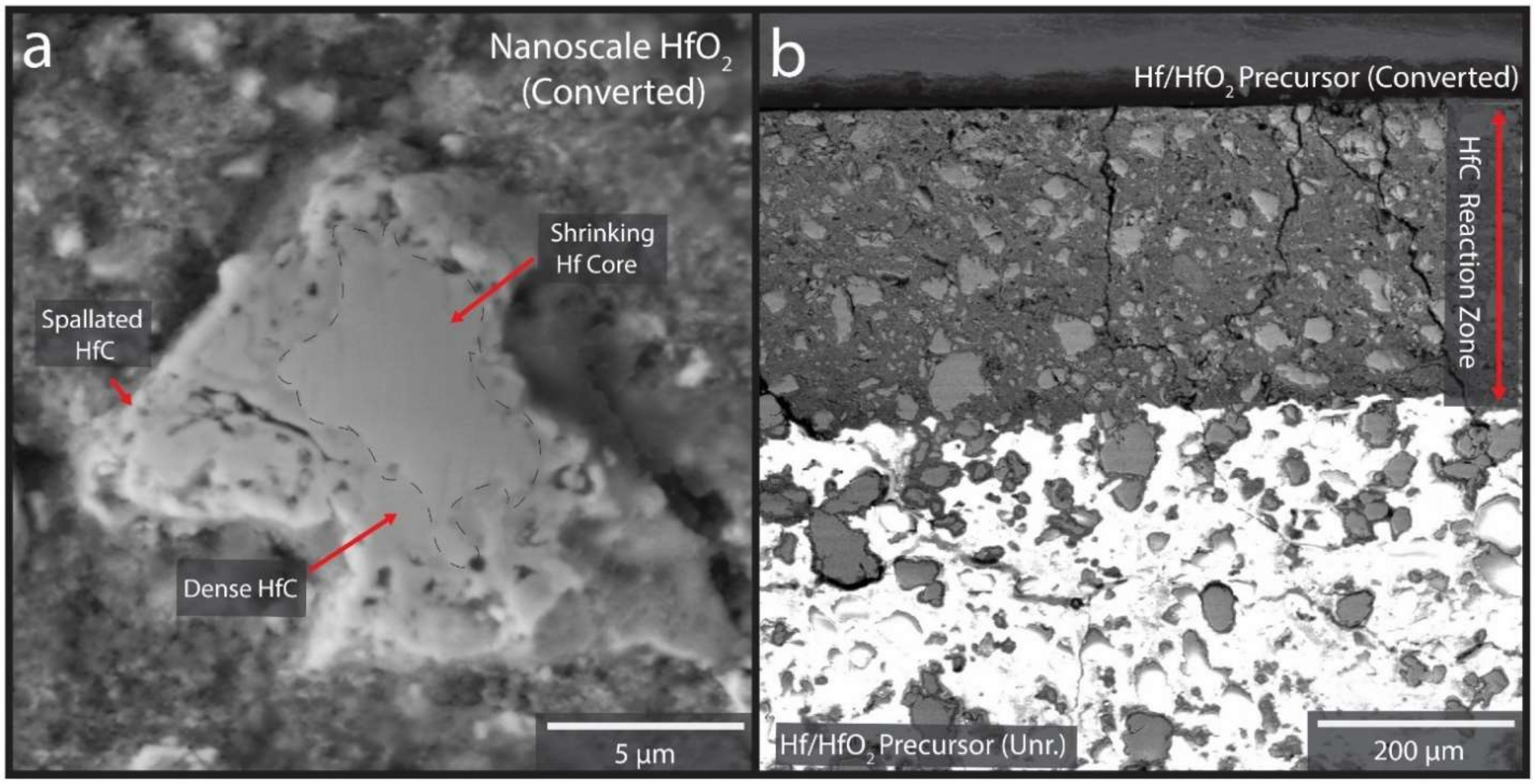

**Figure 11.** SEM micrographs of Hf/$HfO_2$ precursor materials laser processed in $CH_4$ using 4W laser power. (a) Depiction of a Hf particle partially converted to HfC in a matrix of reacted nanoscale $HfO_2$. At the interface of the two precursor materials, spallation is observed on the reacted shell for the 4W proceeding conditions. (b) Cross-section the same Hf/$HfO_2$ precursor material showing depth-dependent conversion. The high SEM accelerating voltage was used to indicate unreacted $HfO_2$ which readily charges due to its low electrical conductivity. Large particles in each image represent converted (or unconverted) Hf.

In summary, near-complete and stoichiometric conversion of metal/metal oxides species was obtained for SLRS syntheses of TiC, and TiN using the high laser energy density processing conditions. Meanwhile, SLRS conversion to ZrC, ZrN, HfC, and HfN produced conversion to oxycarbides, oxynitride, or carbides or nitrides ranging from ~58% $ZrC_{0.88}O_{0.12}$ (from SLRS of Zr/$ZrO_2$ in $CH_4$, 4 W) to >99.9 % yield of $HfC_y$ species containing several substoichiometric and superstoichometric materials phases (from SLRS of Hf/$HfO_2$ in $NH_3$ 5.25-7 W). Results indicate that if metal and or metal-oxide precursors were made compatible for commercial PBF machines, the 100-400 W laser energies capabilities used would be more than sufficient to achieve chemical conversion in reactive atmospheres. Further investigation and tuning of laser processing parameters for a near-stochiometric yield of these UHTC ceramics should be conducted.

**Table 8. X-Ray Composition Analysis Using Rietveld Refinement Modeling: Conversion of Near Net-Shape Precursors in $CH_4$ Using Laser Heating**

| Target Product (Stoichiometric Lattice Parameter) | Solid Phase Precursor (wt% composition) | Target Reaction | Laser Power (W) | Total Carbide or Oxycarbide (wt%) | NaCl-Type $M_xC_y$ or $M_xC_yO_z$, x≈ y+z (wt%) | $M_xC_y$ or $M_xC_yO_z$, Lattice Parameter (Å) | Calc. NaCl-Type Product Stoichiometry | Unr. M (wt%) or C in α-M | Unr. M-O (or oxidation) (wt%) | Calc. Volume Change | Carbon Residue |
|---|---|---|---|---|---|---|---|---|---|---|---|
| **TiC** | Ti/$TiO_2$ (73/27) | 1+2 | 4 | 99.1% | $TiC_y$: 99.1% | 4.327 | ~$TiC_{1.0}$* [55] | ~1.0% | Trace | -1.7% | Trace |
| **4.327 Å** | Ti/$TiO_2$ (73/27) | 1+2 | 5.25 | >99.9 % | $TiC_y$: >99.9 % | 4.327 | ~$TiC_{1.0}$ * [55] | 0% | 0% | -2.1% | Yes |
| **ZrC** | Zr/$ZrO_2$ (79/21) | 3+4 | 4 | 60.8% | $ZrC_yO_z$: 60.8% | 4.691 | $ZrC_{0.95}O_{0.05}$ [71] | 0 | 39.2% | +11.7% | Yes |
| **4.701 Å** | Zr/$ZrO_2$ (79/21) | 3+4 | 5.25 | 90.5% | $ZrC_yO_z$: 90.3% | 4.693 | $ZrC_{0.97}O_{0.03}$ [71] | 0 | 9.7% | +2.4 | Yes |
| **HfC** | Hf/$HfO_2$ (70/30) | 5+6 | 4 | 85.6% | $HfC_y$: 85.6% | 4.640 | ~$HfC_{1.0}$ * [86] | 4.9% | 9.5% | +2.5% | Yes |
| **4.639 Å** | Hf/$HfO_2$ (70/30) | 5+6 | 5.25 | 95.1% | $HfC_y$: 95.1% | 4.641 | ~$HfC_{1.0}$ * [86] | 1.5% | 3.5% | +0.9% | Yes |

*Carbide may contain trace oxygen contaminants and vacancies, but results indicated concentrations were low. †denotes that laser power was increased due to low carbide yield. [32], [34], [55], [56], [59], [71], [84]–[89]

**Table 9. X-Ray Composition Analysis Using Rietveld Refinement Modeling: Conversion of Near Net-Shape Precursors in $NH_3$ Using Laser Heating**

| Target Product (Stoichiometric Lattice Parameter) | Solid Phase Precursor | Target Reaction | Laser Power (W) | Total Nitride and Oxynitride (wt%) | $M_xN_y$, x>y (wt%) | NaCl-Type $M_xN_y$ or $M_xN_yO_z$, x≈ y+z (wt%) | $M_aN_b$, a>b (wt%) | $M_xO_yN_z$ (wt%) | $M_xN_y$ or $M_xN_yO_z$, Lattice Parameter (Å) | Calc. NaCl-Type Product Stoichiometry | Unr. M (wt%) and/or N in α-M | Unr. M-O (or oxidation) (wt%) | Calculated Volume Change |
|---|---|---|---|---|---|---|---|---|---|---|---|---|---|
| TiN | $Ti/TiO_2$ (77/23) | 9+10 | 4 | >99.9 % | - | $TiN_y$: 99.7 | $TiN_{<0.22}$: 0.3% N in α-Ti | - | 4.238 | $TiN_{0.95}$* [67] | - | - | -5.9% |
| 4.241 Å | $Ti/TiO_2$ (77/23) | 9+10 | 5.25 | >99.9 % | - | $TiN_y$: >99.9 | - | - | 4.242 | ~$TiN_{1.0}$* [67] | - | - | -5.2% |
| | $Ti/TiO_2$ (77/23) | 9+10 | 7 | >99.9 % | - | $TiN_y$: >99.9 | - | - | 4.242 | ~$TiN_{1.0}$* [67] | - | - | -5.2% |
| ZrN | $Zr/ZrO_2$ (87/13) | 9+10 | 4 | 95.3% | - | $ZrN_y$: 76.7 | - | $Zr_7N_4O_8$: 18.8% | 4.570 | $ZrN_{0.99}O_{0.01}$ - $ZrN_{0.95}O_{0.05}$ [91] | - | 4.7% | -0.4% |
| 4.568- Å | $Zr/ZrO_2$ (87/13) | 9+10 | 5.25 | 95.2% | - | $ZrN_y$: 86.2 | - | $Zr_7N_4O_8$: 2.1% | 4.574 | $ZrN_{0.75}O_{0.25}$ (est)[91] | - | 4.8% | +1.5% |
| | $Zr/ZrO_2$ (87/13) | 9+10 | 7 | 97.0% | - | $ZrN_y$: 92.7 | - | $Zr_7N_4O_8$: 4.5% | 4.574 | $ZrN_{0.75}O_{0.25}$ (est) [91] | - | 3.0% | -1.8% |
| HfN | $Hf/HfO_2$ (96/4) | 11+12 | 4 | 97.0% | $Hf_3N_4$: 3.2% | $HfN_y$: 34.5% | $Hf_4N_3$: 55.0% $Hf_3N_2$: 1.3% | $Hf_xO_yN_z$ (cub): 1.5% | 4.521 | ~$HfN_{1.0}$ to $Hf_{1.1}$ [92] | $HfN_{<0.29}$: 1.5% Trace Hf (≤1.0%) | 2.0 | +1.6% |
| 4.518 Å | $Hf/HfO_2$ (96/4) | 11+12 | 5.25 | ~98.6% | $Hf_3N_4$: 2.1% | $HfN_y$: 60.2% | $Hf_4N_3$: 34.1% $Hf_3N_2$: 0.7% | $Hf_xO_yN_z$ (cub): trace | 4.520 | ~$HfN_{1.0}$ to $Hf_{1.1}$ [92] | $HfN_{<0.29}$: 2.1% Trace Hf (≤1.4%) | Trace | +0.5% |
| | $Hf/HfO_2$ (96/4) | 11+12 | 7 | >99.9 % | $Hf_3N_4$: 1.3% | $HfN_y$: 64.5% | $Hf_4N_3$: 32.4% $Hf_3N_2$: 0.3% | $Hf_xO_yN_z$ (cub): 0.8% | 4.523 | ~$HfN_{1.0}$ to $Hf_{1.1}$ [92] | $HfN_{<0.29}$: 0.8% Trace | Trace | +1.1% |

*Nitride (or oxynitride species) may contain vacancies

## 3.5 Influence of Reaction Kinetics on Reaction Sintering of UHTC Layers

To this point, much of the analysis of SLRS-UHTCs has been centered on the relationship between non-oxide yields and reaction thermodynamics without considering kinetics. Thorough precursor conversion does not singularly dictate the viability of SLRS for UHTC production and compatibility with powder bed fusion AM. The ability to synthesize numerous refractory ceramics on AM-relevant time scales is only significant for refractory component fabrication if interparticle adhesion can be achieved. Traditional PBF-AM relies on non-reactive binding mechanisms like full/partial melting or sintering that are not generally applicable to UHTCs due to their high melting temperatures and poor response to laser energy deposition. For UHTCs with very low diffusion coefficients, reaction energies associated with synthesis can initiate atomic mobility for interparticle adhesion. This reaction bonding concept may be applied for the simultaneous synthesis and robust layer formation through interparticle adhesion for AM [78].

Reaction bonding during SLRS is based on gas-solid (or liquid) reactivity and a dynamically changing mixture of product reactants and intermediates. Thermal activation is necessary for reactivity, atomic transport, and sintering-induced densification. A generalized approach for kinetic models leads to a rate equation incorporating both thermodynamic variables (reactant gas pressure, temperature) and morphological variables (particle size, porosity, grain size) [96], [97]. The reaction processes and rate between gaseous and solid phases may be governed by either reaction-limited or diffusion-limited according to the thickness of the surface product layer. [98]. The propensity for reaction- or diffusion-limited occur include both chemical kinetics, and processing condition. Further, other considerations like changes in optical absorptivity create dynamic thermal environments and induce microstructural evolution as the product layer is formed.

- Reaction-Limited: If the rate-determining step for the SLRS process is determined by adsorption, desorption or an interfacial reaction, the kinetic rate, dξ/dt (mol $s^{-1}$), is proportional to the surface area:

$$\frac{d\xi}{dt}\text{(surface or interface)} = v_s S \qquad \text{(Eq. 9)}$$

  where $S$ ($m^2$) is the surface area of the reaction zone or particle surface where the rate-determining step occurs and $v_s$ is the rate for the step at the surface or interface according to the Arrhenius dependent rate constants of forward and backward atomic hopping (mol $m^{-2}$ $s^{-1}$). Reaction-limited schemes typically occur when the partial pressure of the gas is low, bulk gas diffusion into the particle-matrix is severely impeded, and/or the sticking coefficient of the gaseous reactant is low [96], [99]. Such conditions cause intergranular diffusion to be faster than adsorption causing the reaction to be slow at the conversion interface.

- Diffusion Limited: If the rate-determining step is diffusion, the rate dξ/dt has a form that relates directly to the shrinking core model where the diffusion of neutral atoms occurs according to diffusive flux $J$ [99].

$$\frac{d\xi}{dt}\text{(diffusion)} = J \bullet S = \frac{D\Delta C}{l_o} G_D S, \qquad \text{(Eq. 10)}$$

  Here $D$ is the diffusion coefficient ($m^2$ $s^{-1}$), $\Delta C$ is the difference in concentration of the diffusing species at the reaction interface, $G_D$ is a dimensionless constant that is a function of particle symmetry. Both $G_D$ and $S$ depend on precursor morphology that varies as a function of time and $J$ is reliant only on thermodynamic processes. If the rate of the reaction is controlled by either external or internal mass transfer, then changes in particle size will affect reaction progression [59]. Morphological changes like alterations in porosity upon conversion will alter this process.

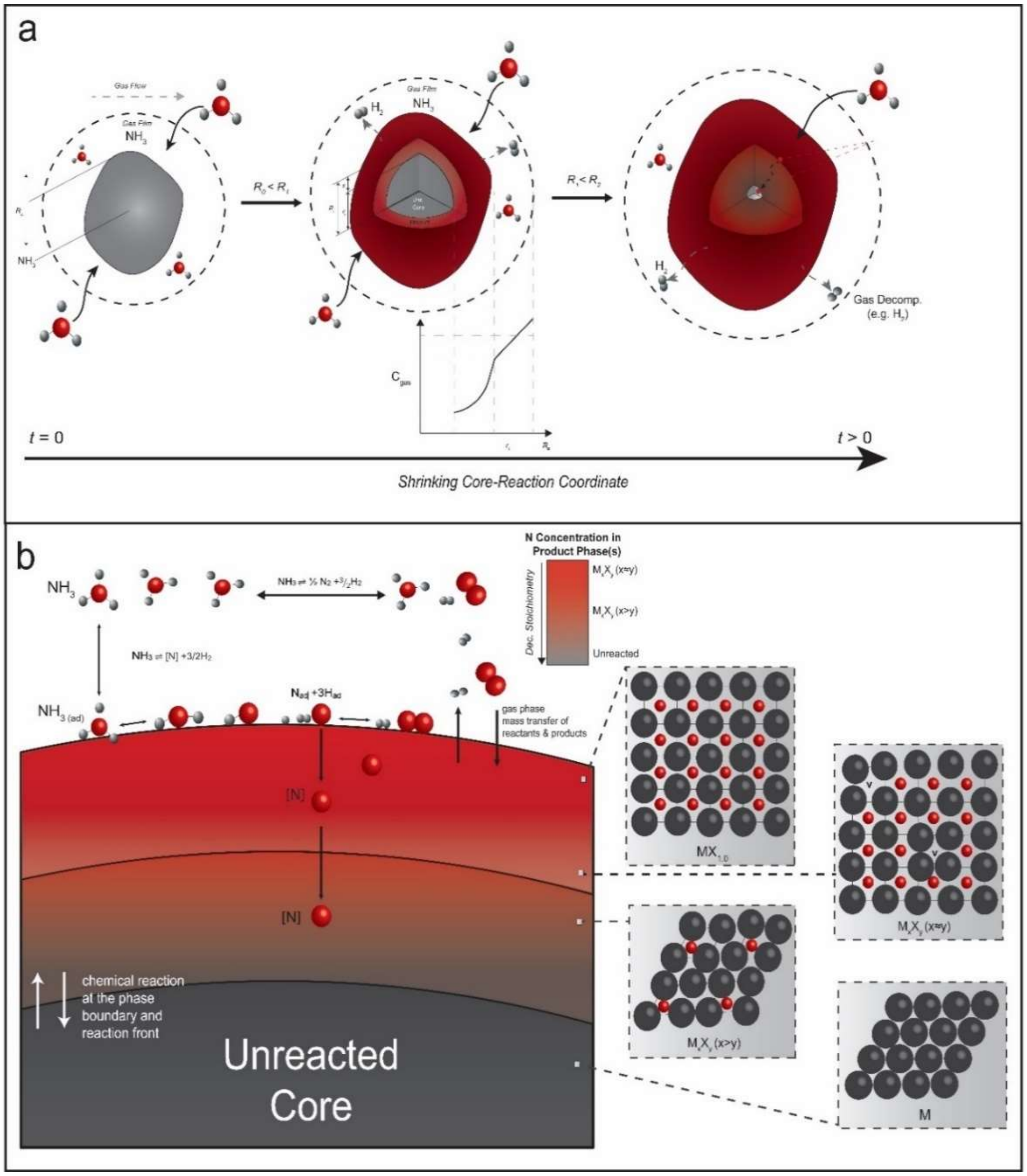

**Figure 12.** (a) Illustration of shrinking core model reaction scheme for particle conversion via gas nitridation of a metal precursor particle with volume expansion (b) Depiction surface adsorption and inter lattice diffusion associated with gas-solid conversion. Depth-dependent conversion and phase compositions are shown.

Several studies on the conversion of dense, transition metal [99] and metal oxide particles [59], [100] to non-oxide ceramics indicate that reactivity can be modeled by a phase composition-dependent shrinking core model as shown in Fig. 13 where diffusion (of reactants or reduction products) through the ceramic product layer is the limiting step. These simple models do not consider complex effects such as the transport of electrons through the product layer, or the transport of carbon or nitrogen ions to the reaction interface by electric field-assisted diffusion, but phenomenologically model reactivity [99]. For dynamic processes, concentration gradients and convective flux may form self-generated electric fields that leave behind charges of the opposite sign. The use of >99.9 vol.% $CH_4$ or $NH_4$ was chosen so the external mass transfer of the gaseous reactant from the bulk gas stream into the powder bed and onto the particle surface would not control the ability to absorb gas molecules onto the precursor surfaces. Under conditions with a fast-flowing pure gas stream, these reactions might be nearly independent of local partial pressure [99]. Therefore, inter-lattice diffusion through an exterior product phase is suggested to be the rate-determining step during much of the rapid laser-induced reaction synthesis process. According to Eq. 10, diffusion through the reacted exterior shell induces non-linear conversion. This type of diffusivity control is qualitatively indicated by the high UHTC yield achieved by low laser energy density but significantly greater energy requirements needed to produce non-oxides with stoichiometries approaching ~1.0.

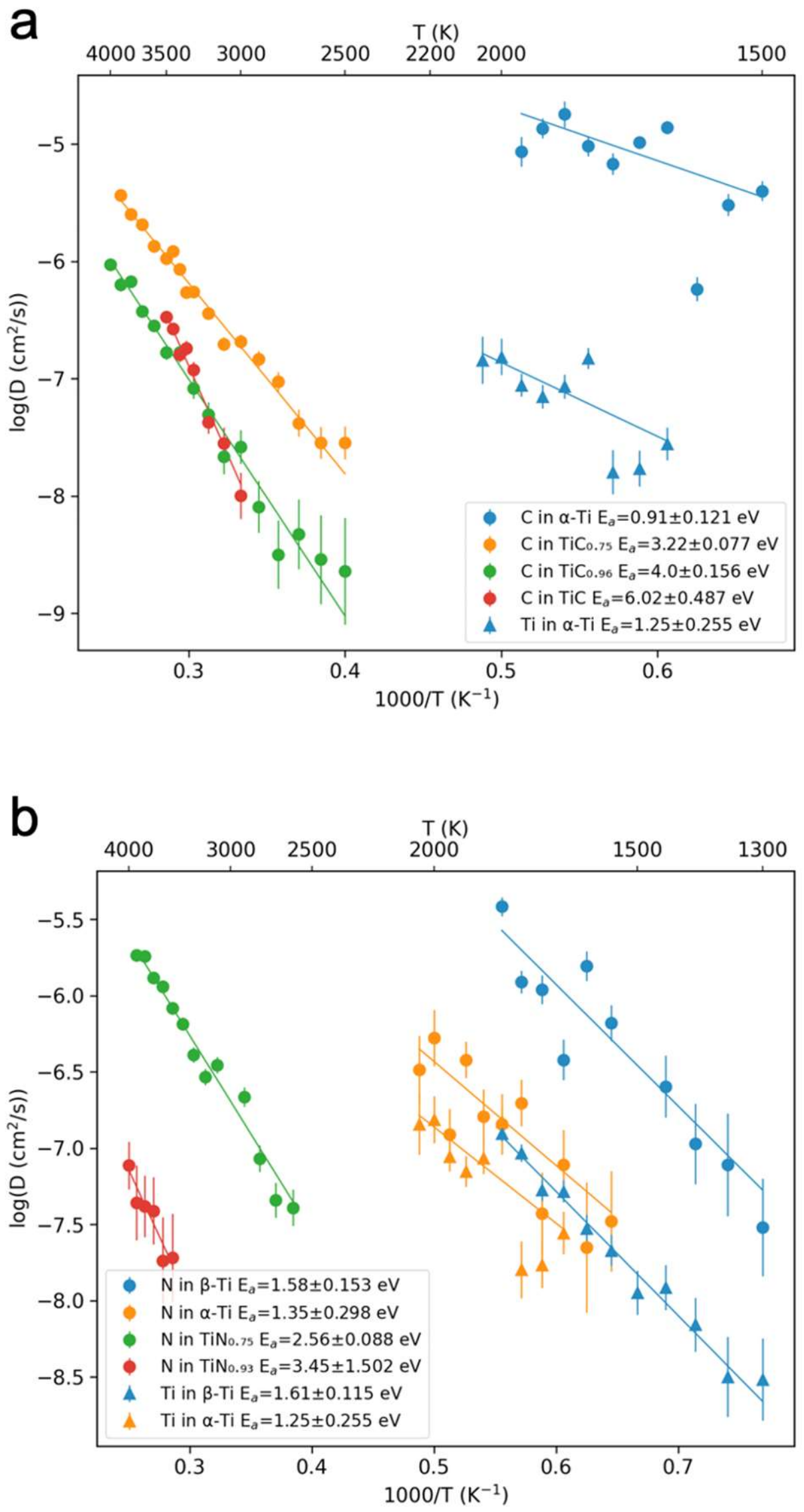

**Figure 13.** Arrhenius plots of diffusion coefficients from LOTF-MD for phases in (a) Ti-C (b) Ti-N. The error bar on each diffusivity data point represents the statistical uncertainty, which is estimated with the method reported by He, X. F., et al.[101] The activation energies shown in the legend are computed from the slope of the fitted line.

Molecular dynamics for the diffusion of nitrogen and carbon within Ti, $TiN_y$ and $TiC_y$ broadly corroborate lattice diffusion rather than adsorption control under the experimental conditions achieved in this work. The Ti-C-N system was chosen as a model system chosen to represent the conversion behavior of group IV UHTC materials. Calculations also qualitatively agree with the prevailing literature where interstitial C forms weaker, less covalent bonds compared to N (or O), which leads to a lower activation energy for diffusion [102]. As such, the atomic radii of C or N atoms do not dictate their ability to diffuse through the interstitial sites of a strained transition metal lattice. Generally, these results provide insight into reaction bonding, chemical conversion, and the ability to form layers of UHTC materials via SLRS-AM.

The self-diffusion of Ti in α-Ti (HCP) or β-Ti (BCC) is predicted to be significantly slower than diffusion of C/N within α-Ti or β-Ti. The diffusion of Ti within α-Ti, β-Ti, TiN, and TiC was too slow for activation energies ($E_a$) to be reliably calculated using active learning molecular dynamics. However, studies indicate $E_a$~6.84-9.84 eV for metal self-diffusion into $TiC_y$ compared to 0.91-6.0 eV for C in $TiC_y$ [84]. Overall the calculated values for C-N diffusion agree with activation energies obtained by others from kinetic measurements of sintering or reaction syntheses of TiC and TiN (Ti-N 2.09-3.86 eV [99]; Ti-C: 3.02-4.85 eV [84], [103]).

Activation energy is approximately proportional to the entropy of melting associated with an elemental diffuser ($\Delta S_m$) and the melting point of the host lattice ($T_m$), $E_a \approx \Delta S_m T_m$ [104]. While the melting point of TiC (3092 °C) is greater than TiN (2930 °C), stronger covalent bonds of Ti-N prevented accurate activation energies to be calculated with reasonable molecular dynamics simulation time due to slow diffusion. Computational methods inform the underlying rate-limiting phenomenon beyond the simple kinetic models by first-principles calculations. Large $E_a$ discrepancies within carbide or nitride phases are suggested to be highly dependent on stoichiometry and particle size effects. Practically, these data suggest that under SLRS heating, C and N may quickly dissolve into interstitial sites of α-Ti ($10^3$-$10^8$ times faster than Ti self-diffusion) to promote the rapid formation of α-Ti-N (0-22 at% N) or α-Ti-C (0-2 at% C). Rapid solubility during low sub-stoichiometric UHTCs progressively impedes further conversion and lattice ductility due to the formation of solute-solvent semi-covalent bonds [67], [102]. Unless activation energies for atomic mobility are overcome by the release of energy due to reactions ($\Delta G_r$) under the local processing conditions the discrete conversion of particles rather than reaction sintering might dominate [105].

Fig. 14 shows how SLRS conversion might impede interparticle adhesion under reaction conditions. The initial conversion of α-Ti may outcompete the atomic diffusion of Ti that might be responsible for pre-reaction sintering. As the interstitial concentration of C or N increases in the $TiC_y$ and $TiN_y$ stoichiometry, continued diffusion is severely impeded (*e.g.* $E_{a\text{-}TiC0.75}$ = 3.22 eV, $E_{a\text{-}TiC1.0}$= 6.02eV; $E_{a\text{-}TiC0.75}$ = 2.59eV, and $E_{a\text{-}TiC1.0}$ was not able to be calculated due to slow atomic diffusion).

## 3.6 Characterization of Preliminary Near Net-Shape Precursor Mixtures

Previous studies on gas-solid reactivity using traditional ceramics processing techniques found that no overall change in macroscopic dimension was observed during reaction synthesis as residual porosity compensated for the reaction-induced volume changes during the nitridation of metallic precursors [99], [106], [107]. However, results from SLRS synthesis of single component precursors and near net-shape metal/metal oxide composition suggest otherwise for photothermal processing regimes. Figs. 7, 9, 11, and 15 indicate that macroscopic layer structure and defect formation is highly dependent on the changes to molar volume, lattice parameter, and local microstructure. SLRS carbidization or nitridation involves processes far from equilibrium such that reaction products and reaction synthesis have vastly different properties compared to conventional ceramics processing [64]. Figure 14 shows the surface morphology of SLRS processed near net-shape metal/metal oxide mixtures. A comparison of phase results in Table 8 and microstructures of $Zr/ZrO_2$ and $Hf/HfO_2$ processed in $CH_4$ (Figure 14) indicate a clear correlation between the degree of stoichiometric conversion and surface defects present in the laser-sintered layers. Specifically, SLRS conversion of 73/27 wt% $Zr/ZrO_2$ to $ZrC_{1.0}$ was incomplete and resulted in oxidation and the production of 53.8 wt% $ZrC_{0.88}O_{0.12}$ and 90.3 wt% $ZrC_{0.97}O_{0.03}$ for the medium (4W) and high power (5.25W) processing conditions respectively. Lattice parameters that were obtained from Rietveld refinement of precursor and reaction products diffractograms and were used to calculate conversion-induced volume change of the $Zr/ZrO_2$ precursor system during conversion in $CH_4$. The calculated volume changes associated with gas-solid conversion ($\Delta V_{calc}$) were determined to be +11.7% and +2.4% respectively for $Zr/ZrO_2$. By contrast, conversion of the 70/30 wt% $Hf/HfO_2$ precursor was associated with net-volume changes +2.5% and +0.9% for the high power and maximum power processing conditions respectively which generally agrees with the theoretical near net shape precursor formulation. Overall, small volume changes are associated with fewer surface defects in the laser processed UHTCs, while large volume changes qualitatively correspond to greater microstructural discontinuity as shown in the SEM micrographs in Figure 14. Similar studies on laser nitridation of bulk metals indicate the presence of microcracks that originated from volume expansion or stress-relief during rapid cooling [27]. While crack behavior can be reduced through the optimization of process parameters such as laser energy, beam dwell time, powder bed preheating etc., unless stoichiometric products are obtained, reactive volume induces stresses that exacerbate crack formation in product materials [74]. The addition of small phase fractions of product phases (e.g. $Ti/TiO_2$ + 10% TiC) might also serve to reduce residual stresses from volumetric changes; however, these phase volumes must be limited to quantities well below the percolation threshold (~16 vol% for spheres), such that reaction bonding is not impeded [108].

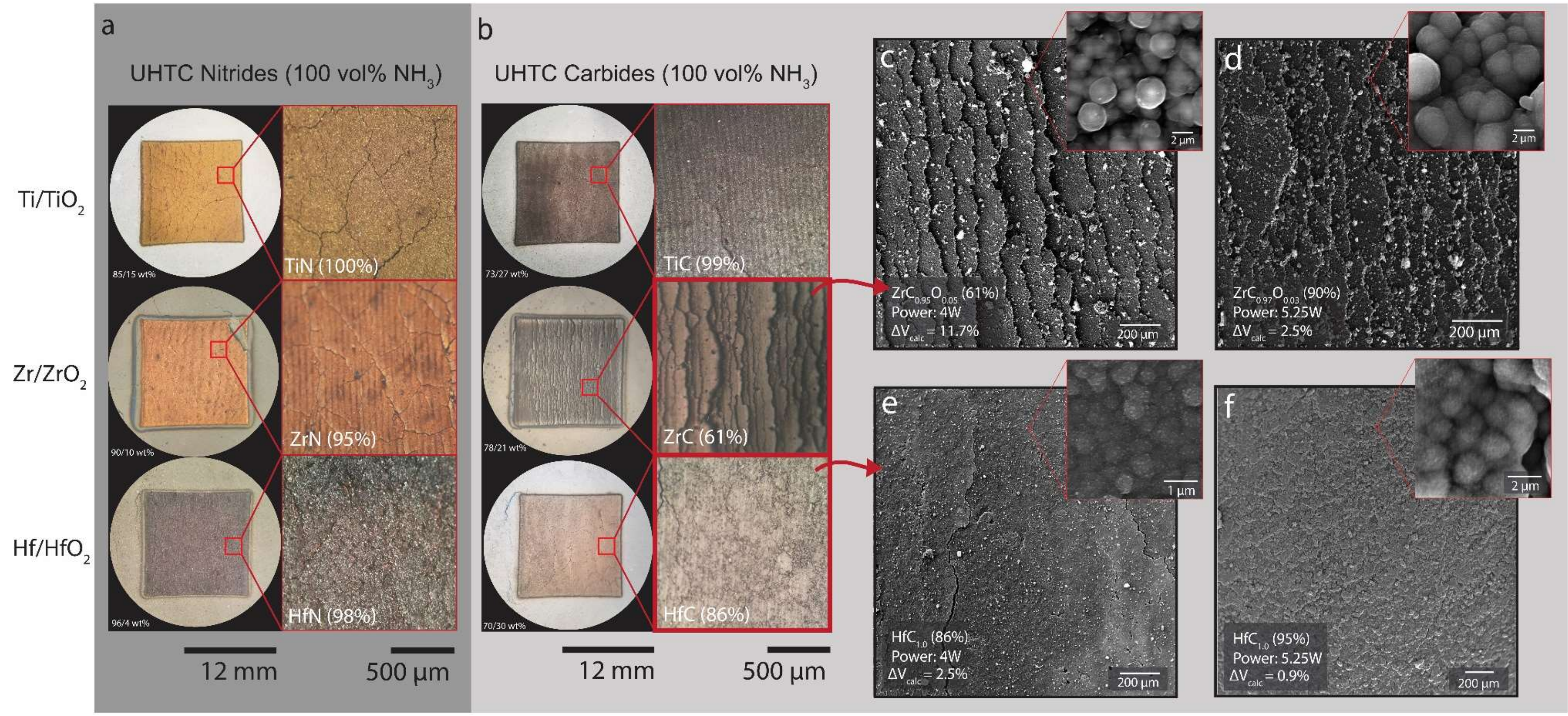

**Figure 14.** Photomicrographs and of metal or metal oxide precursors, laser sintered in $CH_4$ or $NH_3$ using 4W laser power output a-c,e); surface microstructures are depicted for the ZrC and HfC- containing samples via SEM. d and f show the surface microstructure of the same precursor materials when the power is increased to 5.25W. Optical micrographs of samples shown in d and f are depicted in Figs. 9 and 11.

The combined results shown in Fig. 15 indicate that volumetric optimization of precursors appears to be a viable means to improve the microstructure of laser synthesized UHTCs for application with powder bed fusion AM. These observations appeared to be largely independent of particle size and precursor chemistry even if particle size and morphology play a significant role in the thickness of the SLRS product layer [21]. Overall, the macroscopic net-volume change during SLRS processing may be a combination of several factors for laser-assisted gas-solid reactivity of solid particles. For certain oxide precursor mixtures and particle sizes, late-onset reactivity (due to oxide sequential reduction and higher thermodynamic reaction thresholds) allows for higher processing temperatures to be achieved before UHTC materials are formed. Higher processing temperatures might permit the oxide phase to melt (Figure 7d-f) or significantly consolidate (Figure 9c-d) and act as a phase binder in a metal/metal oxide mixture. In this case, the oxide phase could bind UHTC particles formed from metal precursors which do not readily sinter due to their low thermodynamic thresholds for reactivity and slow diffusion. This could occur before the UHTC precipitates out of the melt or converts from the oxide. Densification due to the thermocapillary forces might produce a denser product structure than otherwise achieved by gas-solid reactivity alone. Such densification should be considered in optimizing the precursor for a zero-volume change AM process.

### *3.6.1 Influence of Kinetic Factors on Reaction Bonding and Interparticle Adhesion*

SLRS is a highly non-equilibrium process where precise temperature control is challenging to regulate. Rather, heating rate and particle size selection may have the most profound effect on the mechanical integrity and chemical composition of ceramic product layers. Of the reaction bonded UHTCs laser processed from metal, metal oxide, or metal/metal oxide precursor mixtures only those containing a large wt% of precursor particles with diameters larger than ~15 μm were able to be physically removed from their powder beds. Larger particle sizes were observed to be positively correlated with mechanical integrity in our previous study SLRS processing of $Cr_3C_2$ and CrN. This result may be due in part to the greater sintering depths achieved as beam attenuation is inversely proportional to particle size. However, the *in-situ* synthesis of reaction bonded layers of UHTCs appears to operate through distinctly different mechanisms than other ceramics due to their wide range of interstitial C and N stoichiometries. For example, the Cr-C system is characterized by the occurrence of three intermediate phases in the solid-state: $Cr_{23}C_6$, $Cr_7C_3$, and $Cr_3C_2$. When the transition metal ion (Cr, Mn, Fe, Co, Ni) is smaller than the critical radii, the transition metal carbides form a variety of non-fcc or bcc phases [35]. Each of the carbide phases in the Cr-C system has a relatively narrow homogeneity range of 3 at.% or less compared to group IV transition metal carbides ~50 at% [109]. Previous work [21] has indicated that the intermediate transition metal carbides appeared to form regions of discrete phases ($Cr_{23}C_6$, $Cr_7C_3$, and $Cr_3C_2$) rather than continuous, depth-dependent solid solutions of metal and carbon for Ti-C [109]. Growth of narrow non-oxide stoichiometries at phase interfaces during laser processing might better facilitate transient atomic mobility for reaction bonding compared to the progressive gradient of interstitial carbon or nitrogen saturation of group IV or V transition metal carbides. This mechanistic discrepancy might explain why laser processing for $Cr_3C_2$ formation produced continuous reaction bonded layers that include phases with additional carbon while TiC particles were weakly bonded with the same SLRS conditions (Fig. 15).

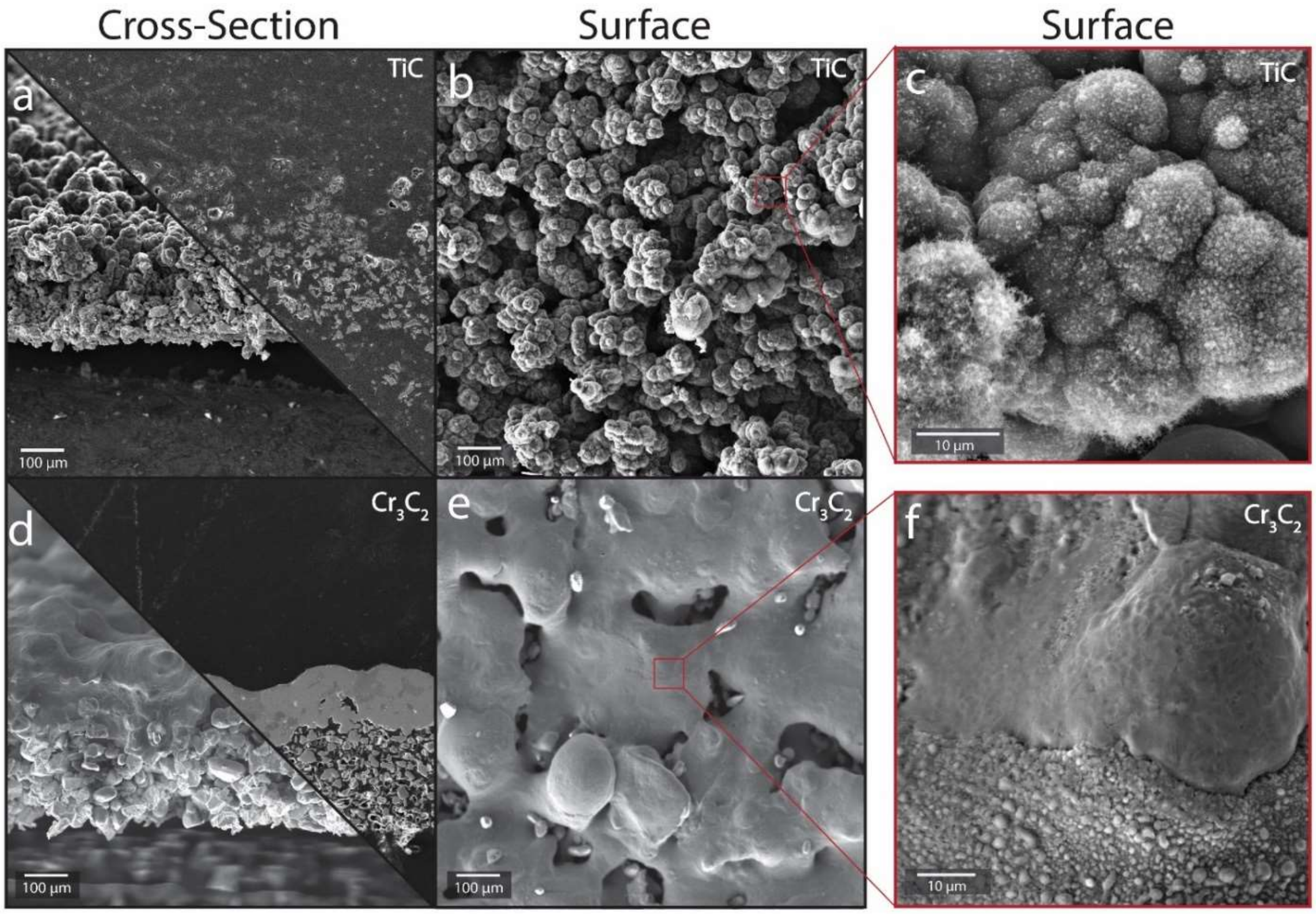

**Figure 15.** SEM micrographs showing the cross-sections and surface morphology of TiC and $Cr_3C_2$ laser processed from (~44 µm) $Ti/TiO_2$ and $Cr/Cr_2O_3$ precursor materials in $CH_4$ using 5.25W laser power.

An ideal SRLS processing trajectory for gas-solid reaction sintering might occur if interparticle adhesion and necking occur before or concurrently with conversion such that the low atomic mobility of the product UHTC does not significantly impede the ability to form a continuous reaction bonding AM layer. Non-oxide synthesis and reaction bonding/densification can occur in sequence or concurrently according to precursor morphology and reaction conditions [110]. For solid-state processing, densification and sintering rate varies with length scale for fixed chemical composition (particle or grain size, *R*), according to $\frac{1}{R^m}$ (Herring's Scaling Law), where *m* is indicative of the transport mechanism: surface or grain boundary diffusion (*m=4*), volume lattice diffusion (*m=3*), vapor transport (*m=2*), plastic or viscous flow (*m=1*) [110], [111]. The time for conversion of a particle becomes proportional to:

$$t \sim \frac{R^m}{D_e}, \qquad \text{(Eq. 11)}$$

where *D* is the Arrhenius dependent diffusion coefficient in corporation activation energies ($E_a$).

Gas-solid reactivity is can be proportional to volume (rate $\propto 1/R^3$) and/or grain boundary diffusion (rate $\propto 1/R^4$) where carbides or nitrides form an exterior non-oxide passivation layer (via the shrinking core model) that slows further diffusion with high interstitial concentrations of C, N. Sintering similarly occurs through volume diffusion and grain boundary diffusion (or surface diffusions at lower temperatures at the beginning of the process) [80]. Volume and grain boundary mechanisms lead to gross densification of a compact (as indicated by the cracking of samples processed in Ar), but surface diffusion does not. SLRS formation of UHTCs by gas-solid reactivity is associated with a temperature-dependent release of free energy upon conversion, $\Delta G_r$ [78], [110], [112]. At high enough temperatures $\Delta G_r$ may be greater than the required activation energy for the diffusion of Ti, C, or N in $TiC_y$ or $TiN_y$. The higher spontaneous reaction thresholds for synthesis from oxide materials (319-2008 ºC) can then favor low-temperature surface diffusion to partially bond particles (especially for small particles where rate $\propto 1/R^4$) during the early stages of the photothermal heating process. By contrast, metal chemistries can begin to spontaneously convert above ~25° so discrete conversion might compete with atomic diffusion that bonds particles during AM layer formation. Low-temperature conversion results in small $\Delta G_r$ and might not provide a significant driving force for enhanced atomic mobility.

The $\Delta G_r$ for TiN and TiC synthesis are shown as a function of temperature in Fig. 17. For example, the reaction, $Ti + CH_4 \xrightarrow{1500K} TiC + 2H_2$ is associated with $\Delta G_r$ ~300kJ/mol (~3.62 eV); which is not sufficient to overcome $E_a$ for the diffusion of C in TiC (6.02 eV).

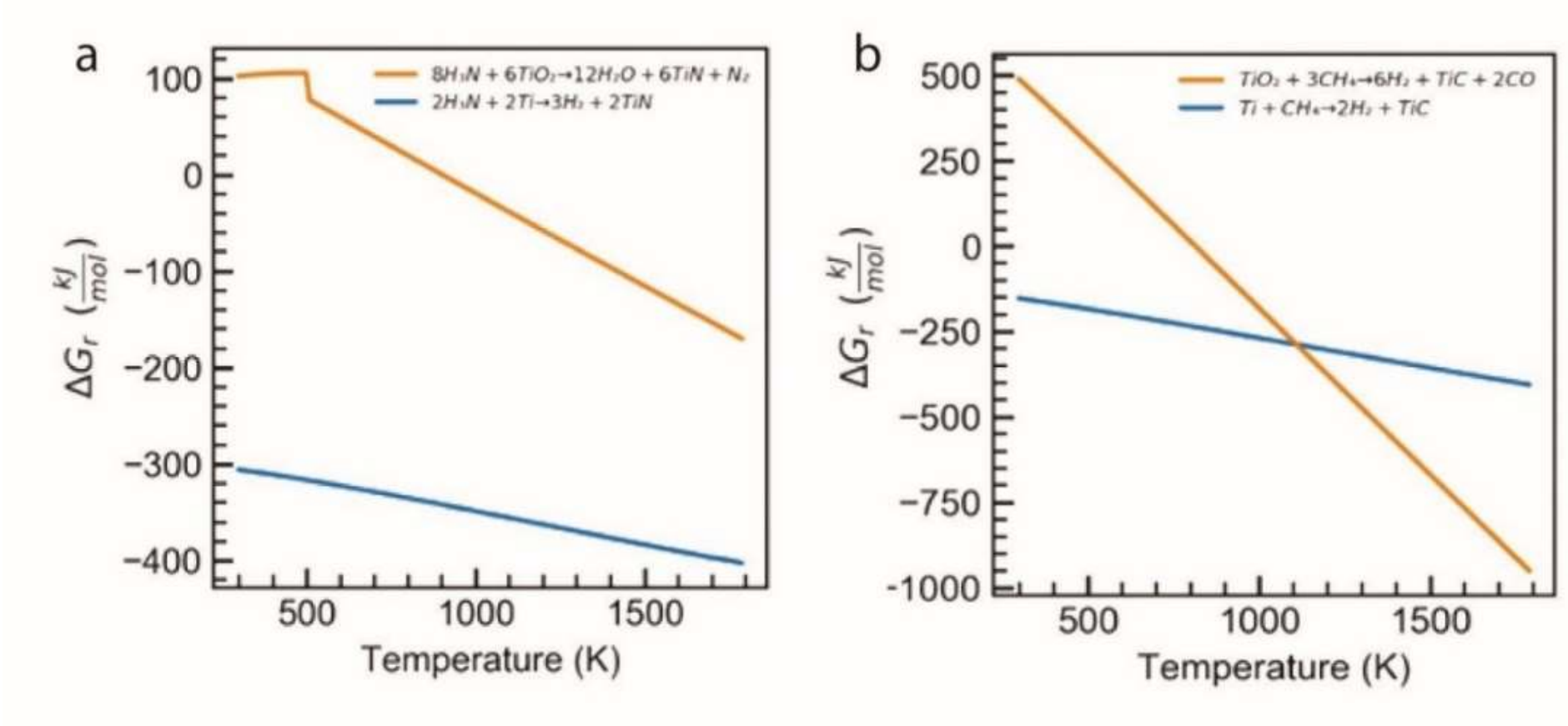

**Figure 16.** Gibbs free energy of reaction diagrams for the formation of TiC and TiN from the reaction of Ti and $TiO_2$ with $CH_4$ and $NH_3$ respectively as a function of temperature.

To best facilitate reaction bonding during AM processing and overcome the high activation energies of UHTCs sintering, the change in free energy for the reaction ($\Delta G_r$) should be leveraged by simultaneously:

1. Achieving high temperatures on time scales faster than gas-phase adsorption so reactivity is limited until conditions favor $\Delta G_r$ (gas-solid reactions):

$$t_{\text{intraparticle-diff}} < t_{\text{heating}} < t_{\text{C/N interparticle diff}} \leq t_{\text{ads}} \quad \text{(Eq. 12)}$$

   The application of high laser energy densities can be used for rapid photothermal energy conversion. If peak temperatures are achieved on timescales shorter than gas-phase adsorption and decomposition of $CH_4$ or $NH_4$, reactivity with large $\Delta G_r$ might be used to aid interstitial atomic transport. Additionally, if grain sizes are small, surface diffusion between particles (rate $\propto 1/R^4$) might dominate over C/N reactivity from volume diffusion (rate $\propto 1/R^3$) and favor interparticle necking during early stages in the reaction.

2. Selecting reaction conditions (gas partial pressure; laser parameters, and particle morphology) so that adsorption and volume diffusion can occur on similar timescales. For gas-solid SLRS, precursor activity is dependent on both surface area and volumes where the rate-limiting step may change throughout the reaction [99]. Increasing particle sizes from nano- to microscale influences the number of adsorption sites per unit precursor mass and diffusive length scales. If particle sizes are large (up to ~50 μm with lower gas-phase availability), the C/N supplied to the particle per unit time will be slowed and regulate the formation of high stoichiometry surface compounds. With sufficient temperature, interstitial atoms might diffuse inward and lead to more uniform conversion without the formation of a dense passivation layer that retards reactivity. Increased particle sizes may have additional benefits including enhanced flowability during powder bed fusion AM and resistance to vaporization under high-laser energy densities.

## 3.7 Conclusion

In this work, SLRS was investigated to produce single-layer UHTC materials representative of reactive additive manufacturing processing. SLRS combines simultaneous carbothermal conversion/reduction (or nitridation) and selective laser processing to bind particles via reaction sintering. Synthesis of several of these refractory materials including HfC has not been previously reported using similar laser-assisted processing techniques. Precursor materials of varied morphology consisting of Ti, $TiO_2$, Zr, $ZrO_2$, and Hf, $HfO_2$ were irradiated in either >99.9 vol% $CH_4$ or $NH_3$ for fabrication of reaction bonded layers of TiC, TiN, ZrC, ZrN, HfC, and HfN respectively. Molecular dynamics calculations were employed to model diffusion kinetic phenomena that inform the experimental results. The combination of experimental and computation methods led to several key conclusions relevant to the maturation of UHTC-AM:

- X-ray diffraction indicated of up to >99.9 wt% total carbides or nitrides could be produced from SLRS in >99.9 vol% $CH_4$ or $NH_3$. Nearly complete stoichiometric conversion was achieved for the reaction of metal precursors to TiC, TiN, and HfC, in part due to their low thermodynamic thresholds for reactivity (>25 ºC), and high optical absorptivity at the processing wavelength (445 nm). By comparison, SLRS processing of metal-oxide precursors produced lower product yields, especially for nitride synthesis which necessitate higher temperatures to overcome thermodynamic thresholds. Higher laser powers were associated with greater total conversion. Generally, conversion of precursor materials indicates that SLRS processing strategies can be almost ubiquitously applied to the synthesis of transition metal carbide and nitrides on AM-relevant time scales if reaction thresholds are satisfied with appropriate laser processing parameters.
- Microstructural evaluation of single-phase metal or metal oxide conversions suggested that conversion-induced volumetric expansion (metal) and volumetric contraction (metal oxide) led to severe internal stresses that prevented appreciable mechanical integrity from being obtained. Internal stresses in product materials led to microstructural defects, fracture, and layer bucking. For chemistries where reaction-induced expansion occurred, concave layers bowed off their powder beds; when conversion induced volumetric reduction, layers formed convex hulls and mudcrack-like microstructure. These results were broadly independent of chemistry or particle size and were exacerbated with higher non-oxide yields.
- Near net-shape metal/metal oxide composite precursors were employed to mitigate these single-component volume changes. Results indicate that volumetric optimization may be viable as a means to improve the microstructure of laser-synthesized UHTCs. Rietveld refinement was used to determine the lattice parameters of precursors and products to estimate conversion-induced volume changes. Precursor mixtures might be tailored for a desired UHTC stoichiometry given that defect species influence the product microstructure. Even so, crack-free UHTC product layers were able to be produced with total reaction-induced volume changes as low as +0.9%.

Optimization of precursor morphology and processing parameters might be used to further improve these results.

- Reaction bonding combines chemical reactivity and inter/intraparticle diffusion, however, these phenomena are not entirely interdependent. Under certain processing regimes, chemical conversion can occur before sufficient transport for interparticle is achieved. The rate at which conversion occurs relative to atomic diffusion is important for SLRS production of dense mechanically robust ceramics. Molecular dynamics calculations indicate that UHTC formation significantly reduces atomic mobility that might lead to particle bonding. Progressive (diffusion limiting) conversion to rocksalt-interstitial carbides or nitrides might be circumvented by utilizing higher energy densities to heat particles above reaction temperatures before gas-phase adsorption occurs. This way, precipitation from the melt and/or large $\Delta G_r$ upon reactivity can be leveraged to help overcome activation energies required for atomic transport.

More detailed, chemistry-specific investigations into processing parameters and precursor optimization should be conducted to advance SLRS methodologies as a viable means for AM-UHTC production. The methods presented here are promising for the fabrication of unique ceramic chemistries that are generally incompatible with current AM processing strategies.

## 4. Declaration of Competing Interest

Funding for the study described in this publication was provided by the Office of Naval Research. Under a license agreement between Synteris LLC and Johns Hopkins University, A. Peters and the University are entitled to royalty distributions related to the technology described in the study discussed in this publication. A. Peters is a founder of and holds equity in Synteris LLC. He also serves as the Chief Technical Officer and holds a board seat to Synteris LLC. The results of the study discussed in this publication could affect the value of Synteris. This arrangement is pending review and approval by Johns Hopkins University in accordance with its conflict of interest policies. D. Nagle, D. Zhang and M. Brupbacher are entitled to royalty distributions related to the technology described in the study discussed in this publication. D. Nagle, D. Zhang, and M. Brupbacher are not affiliated with Synteris LLC. This arrangement has been reviewed by the Johns Hopkins University in accordance with its conflict of interest policies.

## 5. Acknowledgments

The authors gratefully acknowledge funding provided by the Office of Naval Research, Nanomaterials Program Office, under contract N00014-16-1-2460 in partial support of this research. We would also like to thank the Johns Hopkins Applied Physics Laboratory Graduate Fellowship Committee for graduate student support and Jarod Gagnon for useful discussions on this project.

## 6. References

[1] T. H. Squire and J. Marschall, "Material property requirements for analysis and design of UHTC components in hypersonic applications," *J Eur Ceram Soc*, vol. 30, no. 11, pp. 2239–2251, 2010, doi: 10.1016/j.jeurceramsoc.2010.01.026.

[2] W. G. Fahrenholtz and G. E. Hilmas, "Ultra-high temperature ceramics: Materials for extreme environments," *Scr Mater*, vol. 129, pp. 94–99, 2017, doi: 10.1016/j.scriptamat.2016.10.018.

[3] W. Fahrenholtz, Wuchina. E, W. Lee, and Y. Zhou, *Ultra High Temperature Ceramics Materials for Extreme Environment Applications*. 2014.

[4] T. A. Parthasarathy, M. D. Petry, M. K. Cinibulk, T. Mathur, and M. R. Gruber, "Thermal and oxidation response of UHTC leading edge samples exposed to simulated hypersonic flight conditions," *Journal of the American Ceramic Society*, vol. 96, no. 3, pp. 907–915, 2013, doi: 10.1111/jace.12180.

[5] E. Feilden, D. Glymond, E. Saiz, and L. Vandeperre, "High temperature strength of an ultra high temperature ceramic produced by additive manufacturing," *Ceram Int*, vol. 45, no. 15, pp. 18210–18214, 2019, doi: 10.1016/j.ceramint.2019.05.032.

[6] D. Piccione, "Ballistic Evaluation of Rolled Homogenious Steel Armor with Tungsten Carbide and Titanium Carbide Facing," 1960.

[7] H. O. Pierson, "Handbook of Carbides and Nitrides," *Handbook of Refractory Carbides and Nitrides*, pp. 100–117, 1996, doi: 10.1016/b978-081551392-6.50007-6.

[8] S. V. Ushakov, A. Navrotsky, Q. J. Hong, and A. van de Walle, "Carbides and nitrides of zirconium and hafnium (Supplementary)," *Materials*, vol. 12, no. 7, pp. 1–23, 2019, doi: 10.3390/ma12172728.

[9] S. V. Ushakov, A. Navrotsky, Q. J. Hong, and A. van de Walle, "Carbides and nitrides of zirconium and hafnium," *Materials*, vol. 12, no. 7, 2019, doi: 10.3390/ma12172728.

[10] Z. Gu *et al.*, "Identification and thermodynamic mechanism of the phase transition in hafnium nitride films," *Acta Mater*, vol. 90, pp. 59–68, 2015, doi: 10.1016/j.actamat.2015.02.026.

[11] A. Ul-Hamid, "Microstructure, properties and applications of Zr-carbide, Zr-nitride and Zr-carbonitride coatings: a review," *Mater Adv*, vol. 1, no. 5, pp. 1012–1037, 2020, doi: 10.1039/d0ma00233j.

[12] A. Nino, T. Hirabara, S. Sugiyama, and H. Taimatsu, "Preparation and characterization of tantalum carbide (TaC) ceramics," *Int J Refract Metals Hard Mater*, vol. 52, pp. 203–208, 2015, doi: 10.1016/j.ijrmhm.2015.06.015.

[13] W. Lengauer, "Transition Metal Carbides, Nitrides, and Carbonitrides," in *Handbook of Ceramic Hard Materials*, no. October, Weinheim, Germany: Wiley-VCH Verlag GmbH, 2008, pp. 202–252. doi: 10.1002/9783527618217.ch7.

[14] P. Goursat and S. Foucaud, "Non-oxide Ceramics," in *Ceramic Materials: Processes, Properties, and Applications*, 2007.

[15] C. W. Watson, *Nuclear Rockets: High-Performance Propulsion for Mars*, no. May. Los Alamos, New Mexico: Los Alamos National Lab, 1994.

[16] Q. Jia and D. Gu, "Selective laser melting additive manufacturing of Inconel 718 superalloy parts: Densification, microstructure and properties," *J Alloys Compd*, vol. 585, pp. 713–721, 2014, doi: 10.1016/j.jallcom.2013.09.171.

[17] A. Iveković *et al.*, "Selective laser melting of tungsten and tungsten alloys," *Int J Refract Metals Hard Mater*, vol. 72, no. November 2017, pp. 27–32, 2018, doi: 10.1016/j.ijrmhm.2017.12.005.

[18] M. C. Leu, S. Pattnaik, and G. E. Hilmas, "Investigation of laser sintering for freeform fabrication of zirconium diboride parts," *Virtual Phys Prototyp*, vol. 7, no. 1, pp. 25–36, Mar. 2012, doi: 10.1080/17452759.2012.666119.

[19] S. Meyers, "Additive Manufacturing of Technical Ceramics Laser Sintering of Alumina and Silicon Carbide," KU Leuven, 2019.

[20] T. Chartier, "Ceramic Forming Processes," in *Ceramic Materials*, P. Boch and J.-C. N. Niepce, Eds. London, UK: ISTE, 2007, pp. 123–197. doi: 10.1002/9780470612415.ch5.

[21] A. B. Peters *et al.*, "Selective laser sintering in reactive atmospheres : Towards in-situ synthesis of net-shaped carbide and nitride ceramics," *Addit Manuf*, no. January, 2021.

[22] D. Höche, J. Kaspar, and P. Schaaf, "Laser nitriding and carburization of materials," *Laser Surface Engineering: Processes and Applications*, pp. 33–58, 2015, doi: 10.1016/B978-1-78242-074-3.00002-7.

[23] I. Ursu *et al.*, "Titanium and zirconium nitridation under the action of microsecond pulsed TEA CO2 laser radiation in technical nitrogen," *J Phys D Appl Phys*, vol. 18, no. 8, pp. 1693–1700, 1985, doi: 10.1088/0022-3727/18/8/031.

[24] I. Ursu, I. N. Mihailescu, A. M. Prokhorov, V. I. Konov, V. N. Tokarev, and S. A. Uglov, "On the mechanism of surface compound formation by powerful microsecond pulsed TEA CO2 laser irradiation in technical nitrogen," *J Phys D Appl Phys*, vol. 18, no. 12, pp. 2547–2555, 1985, doi: 10.1088/0022-3727/18/12/023.

[25] I. Ursu *et al.*, "Surface nitridation of zirconium and hafnium by powerful cw CO_2 laser irradiation in air," *Appl Opt*, vol. 25, no. 16, p. 2720, 1986, doi: 10.1364/ao.25.002720.

[26] I. Ursu *et al.*, "Multi-pulse laser nitridation of titanium, zirconium and hafnium in a nitrogen atmosphere containing oxygen," *J Phys D Appl Phys*, vol. 20, no. 11, pp. 1519–1524, 1987, doi: 10.1088/0022-3727/20/11/025.

[27] J. D. Wu, C. Z. Wu, X. X. Zhong, Z. M. Song, and P. M. Li, "Surface nitridation of transition metals by pulsed laser irradiation in gaseous nitrogen," *Surf Coat Technol*, vol. 96, no. 2–3, pp. 330–336, 1997, doi: 10.1016/S0257-8972(97)00267-3.

[28] J. X. Liu, Y. M. Kan, and G. J. Zhang, "Synthesis of ultra-fine hafnium carbide powder and its pressureless sintering," *Journal of the American Ceramic Society*, vol. 93, no. 4, pp. 980–986, 2010, doi: 10.1111/j.1551-2916.2009.03531.x.

[29] A. B. Peters *et al.*, "A Reaction Synthesis Approach to Additively Manufacture Net-Shape Chromium Carbide and Non-Oxide Refractory Ceramics," *Addit Manuf*, vol. 34, 2020, doi: 10.1016/j.addma.2020.101186.

[30] A. B. Peters *et al.*, "Isovolumetric synthesis of chromium carbide for selective laser reaction sintering (SLRS)," *Int J Refract Metals Hard Mater*, vol. 83, no. April, p. 104967, Sep. 2019, doi: 10.1016/j.ijrmhm.2019.05.013.

[31] I. Gibson and D. Rosen, *Additive Manufacturing Technologies*, Second. Springer, 2015.

[32] F. Réjasse, O. Rapaud, G. Trolliard, O. Masson, and A. Maître, "Experimental investigation and thermodynamic evaluation of the C–Hf–O ternary system," *Journal of the American Ceramic Society*, vol. 100, no. 8, pp. 3757–3770, 2017, doi: 10.1111/jace.14901.

[33] Y. Lo Vegard, "Die Konstitution der Mischkristalle und die Raumffillung der Atome," *Zeitschrift fur Physik*, 1921.

[34] C. H. Dahm, "Chemical Vapor Deposition of Titanium Oxycarbide A thesis presented for the degree of Master of Science," Lund University, 2020.

[35] "Carbides," *Lumen Courses*.

[36] C. Wang, K. Aoyagi, P. Wisesa, and T. Mueller, "Lithium Ion Conduction in Cathode Coating Materials from On-the-Fly Machine Learning," *Chemistry of Materials*, vol. 32, no. 9, pp. 3741–3752, 2020, doi: 10.1021/acs.chemmater.9b04663.

[37] A. V Shapeev, "Moment Tensor Potentials: A Class of Systematically Improvable Interatomic Potentials," *Multiscale Modeling & Simulation*, vol. 14, no. 3, pp. 1153–1173, Jan. 2016, doi: 10.1137/15M1054183.

[38] A. V Shapeev, "Moment Tensor Potentials: A Class of Systematically Improvable Interatomic Potentials," *Multiscale Modeling & Simulation*, vol. 14, no. 3, pp. 1153–1173, Jan. 2016, doi: 10.1137/15M1054183.

[39] E. V. Podryabinkin and A. V. Shapeev, "Active learning of linearly parametrized interatomic potentials," *Comput Mater Sci*, vol. 140, pp. 171–180, 2017, doi: 10.1016/j.commatsci.2017.08.031.

[40] S. Plimpton, "Fast parallel algorithms for short-range molecular dynamics," *Journal of Computational Physics*, vol. 117, no. 1. pp. 1–19, 1995. doi: 10.1006/jcph.1995.1039.

[41] B. R. Birmingham and H. L. Marcus, "Solid Freeform Fabrication of Silicon Nitride Shapes by Selective Laser Reaction Sintering ( SLRS )," pp. 389–396, 1995.

[42] B. Almangour, *Additive manufacturing of emerging materials*. Springer International, 2018. doi: 10.1007/978-3-319-91713-9.

[43] K. G. Prashanth, S. Scudino, T. Maity, J. Das, and J. Eckert, "Is the energy density a reliable parameter for materials synthesis by selective laser melting?," *Mater Res Lett*, vol. 5, no. 6, pp. 386–390, 2017, doi: 10.1080/21663831.2017.1299808.

[44] Y. P. Kathuria, "Laser surface nitriding of yttria stabilized tetragonal zirconia," *Surf Coat Technol*, vol. 201, no. 12, pp. 5865–5869, 2007, doi: 10.1016/j.surfcoat.2006.10.041.

[45] K. T. Jacob, R. Verma, and R. M. Mallya, "Nitride synthesis using ammonia and hydrazine - A thermodynamic panorama," *J Mater Sci*, vol. 37, no. 20, pp. 4465–4472, 2002, doi: 10.1023/A:1020649913206.

[46] P. Schaaf, "Laser Nitriding in Materials," *Prog Mater Sci*, vol. 47, pp. 1–161, 2002.

[47] J. T. Slycke, E. J. Mittemeijer, and M. A. J. Somers, *Thermodynamics and kinetics of gas and gas-solid reactions*. Woodhead Publishing Limited, 2015. doi: 10.1533/9780857096524.1.3.

[48] E. Giorgi, S. Grasso, E. Zapata-Solvas, and W. E. Lee, "Reactive carbothermal reduction of ZrC and ZrOC using Spark Plasma Sintering," *Advances in Applied Ceramics*, vol. 117, no. sup1, pp. s34–s47, 2018, doi: 10.1080/17436753.2018.1510817.

[49] R. W. Harrison and W. E. Lee, "Processing and properties of ZrC, ZrN and ZrCN ceramics: a review," *Advances in Applied Ceramics*, vol. 115, no. 5, pp. 294–307, 2016, doi: 10.1179/1743676115Y.0000000061.

[50] I. N. Mihailescu *et al.*, "Direct carbide synthesis by multipulse excimer laser treatment of Ti samples in ambient CH4 gas at superatmospheric pressure," vol. 5286, no. August 1993, 1998.

[51] C. Boulmer-Leborgne *et al.*, "Direct carbidation of titanium as a result of multipulse UV-laser irradiation of titanium samples in an ambient methane gas," *Appl Surf Sci*, vol. 54, no. C, pp. 349–352, 1992, doi: 10.1016/0169-4332(92)90069-A.

[52] T. Kühnle and K. Partes, "In-situ Formation of Titanium Boride and Titanium Carbide by Selective Laser Melting," vol. 39, pp. 432–438, 2012, doi: 10.1016/j.phpro.2012.10.058.

[53] Y. Suda, H. Kawasaki, K. Doi, J. Nanba, and T. Ohshima, "Preparation of crystalline TiC thin films grown by pulsed Nd:YAG laser deposition using Ti target in methane gas," *Mater Charact*, vol. 48, no. 2–3, pp. 221–228, 2002, doi: 10.1016/S1044-5803(02)00243-7.

[54] Y. Leconte, H. Maskrot, L. Combemale, N. Herlin-Boime, and C. Reynaud, "Application of the laser pyrolysis to the synthesis of SiC, TiC and ZrC pre-ceramics nanopowders," *J Anal Appl Pyrolysis*, vol. 79, no. 1-2 SPEC. ISS., pp. 465–470, 2007, doi: 10.1016/j.jaap.2006.11.009.

[55] D. N. Miller *et al.*, "Studies on the crystal structure, magnetic and conductivity properties of titanium oxycarbide solid solution (TiO1-xCx)," *J Mater Chem A Mater*, vol. 4, no. 15, pp. 5730–5736, 2016, doi: 10.1039/c6ta00042h.

[56] H. Nakayama, K. Ozaki, T. Nabeta, and Y. Nakajima, "Composition dependence of lattice parameter, thermal and electrical properties in ZrCx compounds," *Mater Trans*, vol. 58, no. 6, pp. 852–856, 2017, doi: 10.2320/matertrans.M2016283.

[57] T. S. Rajaraman, S. P. Parikh, and V. G. Gandhi, "Black TiO2: A review of its properties and conflicting trends," *Chemical Engineering Journal*, vol. 389, no. July 2019, p. 123918, Jun. 2020, doi: 10.1016/j.cej.2019.123918.

[58] N. Wetchakun, B. Incessungvorn, K. Wetchakun, and S. Phanichphant, "Influence of calcination temperature on anatase to rutile phase transformation in TiO2 nanoparticles synthesized by the modified sol–gel method," *Mater Lett*, vol. 82, pp. 195–198, Sep. 2012, doi: 10.1016/j.matlet.2012.05.092.

[59] G. Zhang, "Reduction of Rutile and Ilmenite by Methane- Hydrogen Gas Mixture," The University ofNew South Wales Faculty, 2000.

[60] Z. Lv and J. Dang, "Mathematical modeling of the reaction of metal oxides with methane," *RSC Adv*, vol. 10, no. 19, pp. 11233–11243, 2020, doi: 10.1039/c9ra09418k.

[61] L. M. Berger, W. Gruner, E. Langholf, and S. Stolle, "On the mechanism of carbothermal reduction processes of TiO2 and ZrO2," *Int J Refract Metals Hard Mater*, vol. 17, no. 1, pp. 235–243, 1999, doi: 10.1016/S0263-4368(98)00077-8.

[62] D. Höche, M. Shinn, J. Kaspar, G. Rapin, and P. Schaaf, "Laser pulse structure dependent texture of FEL synthesized TiNx coatings," *J Phys D Appl Phys*, vol. 40, no. 3, pp. 818–825, 2007, doi: 10.1088/0022-3727/40/3/019.

[63] E. D'Anna *et al.*, "Surface nitridation by multipulse excimer laser irradiation," *Prog Surf Sci*, vol. 35, no. 1–4, pp. 129–142, 1990, doi: 10.1016/0079-6816(90)90031-E.

[64] Tatiana and V. Khatko, "Laser synthesis of nitrides on the surface of refractory metals immersed in liquid nitrogen," *Appl Surf Sci*, vol. 254, no. 4, pp. 961–965, 2007, doi: 10.1016/j.apsusc.2007.08.080.

[65] H. Shen and L. Wang, "Formation, tribological and corrosion properties of thicker Ti-N layer produced by plasma nitriding of titanium in a N2-NH3 mixture gas," *Surf Coat Technol*, vol. 393, no. February, p. 125846, 2020, doi: 10.1016/j.surfcoat.2020.125846.

[66] H. Eskandari Sabzi, "Powder bed fusion additive layer manufacturing of titanium alloys," *Materials Science and Technology (United Kingdom)*, vol. 35, no. 8, pp. 875–890, 2019, doi: 10.1080/02670836.2019.1602974.

[67] H. A. Wriedt and J. L. Murray, "The N-Ti (Nitrogen-Titanium) system," *Bulletin of Alloy Phase Diagrams*, vol. 8, no. 4, pp. 378–388, 1987, doi: 10.1007/BF02869274.

[68] H. Kageyama, "Macroporous titanium compound monolith and method for manufacturing same," EP 2 816 012 A1, 2014

[69] S. Naim Katea, L. Riekehr, and G. Westin, "Synthesis of nano-phase ZrC by carbothermal reduction using a ZrO2–carbon nano-composite," *J Eur Ceram Soc*, vol. 41, no. 1, pp. 62–72, 2021, doi: 10.1016/j.jeurceramsoc.2020.03.055.

[70] L. Combemale, Y. Leconte, X. Portier, N. Herlin-boime, and C. Reynaud, "Synthesis of nanosized zirconium carbide by laser pyrolysis route," vol. 483, pp. 468–472, 2009, doi: 10.1016/j.jallcom.2008.07.159.

[71] I. L. Shabalin, "Zirconium Monocarbide," in *Ultra-High Temperature Materials II*, Dordrecht: Springer Netherlands, 2019, pp. 423–675. doi: 10.1007/978-94-024-1302-1_5.

[72] I. Ursu *et al.*, "Synthesis of sheet conductive layers on the surface of some insulator ceramics (TiO2, ZrO2, HfO2) by multipulse CO2-laser irradiation in an ammonia atmosphere," *J Appl Phys*, vol. 66, no. 8, pp. 3682–3687, 1989, doi: 10.1063/1.344080.

[73] I. Ursu, I. N. Mihailescu, L. Nanu, A. M. Prokhorov, V. I. Konov, and V. G. Ralchenko, "Nitrification of zirconium by cw CO2 laser irradiation in ambient atmosphere," *Appl Phys Lett*, vol. 46, no. 2, pp. 110–112, 1985, doi: 10.1063/1.95702.

[74] A. Ul-Hamid, "The effect of deposition conditions on the properties of Zr-carbide, Zr-nitride and Zr-carbonitride coatings – a review," *Mater Adv*, vol. 1, no. 5, pp. 988–1011, 2020, doi: 10.1039/d0ma00232a.

[75] S. Niyomsoan, W. Grant, D. L. Olson, and B. Mishra, "Variation of color in titanium and zirconium nitride decorative thin films," *Thin Solid Films*, vol. 415, no. 1–2, pp. 187–194, 2002, doi: 10.1016/S0040-6090(02)00530-8.

[76] B. Qian and Z. Shen, "Laser sintering of ceramics," *Journal of Asian Ceramic Societies*, vol. 1, no. 4, pp. 315–321, 2013, doi: 10.1016/j.jascer.2013.08.004.

[77] J. Szekely and J. W. Evans, "A structural model for gas-solid reactions with a moving boundary," *Chem Eng Sci*, vol. 25, pp. 1091–1107, 1970.

[78] H. Y. Sohn and J. Szekely, "A structural model for gas-solid reactions with a moving boundary-IV. Langmuir-Hinshelwood kinetics," *Chem Eng Sci*, vol. 28, no. 2, pp. 1169–1177, 1973.

[79] H. Y. Sohn and J. Szekely, "A structural model for gas-solid reactions with a moving boundary — III A general dimensionless representation of the ...," *Chem Eng Sci*, vol. 27, pp. 763–778, 1972.

[80] J. Szekely, *Gas-Solid Reactions*. Elsevier Science, 2012.

[81] M. A. Qayyum and D. A. Reeve, "Hydrogenation of carbon to methane in reduced sponge iron, chromium and ferrochromium," *Carbon N Y*, vol. 14, no. 4, pp. 199–202, 1976, doi: 10.1016/0008-6223(76)90107-X.

[82] L. Haferkamp, A. Spierings, M. Rusch, D. Jermann, M. A. Spurek, and K. Wegener, "Effect of Particle size of monomodal 316L powder on powder layer density in powder bed fusion," *Progress in Additive Manufacturing*, 2020, doi: 10.1007/s40964-020-00152-4.

[83] J. Tang *et al.*, "Martensitic phase transformation of isolated HfO2, ZrO 2, and HfxZr1-xO2 ($0 < x < 1$) nanocrystals," *Adv Funct Mater*, vol. 15, no. 10, pp. 1595–1602, 2005, doi: 10.1002/adfm.200500050.

[84] I. L. Shabalin, "Titanium Monocarbide," in *Ultra-High Temperature Materials III*, Dordrecht: Springer Netherlands, 2020, pp. 11–514. doi: 10.1007/978-94-024-2039-5_2.

[85] P. Barnier and F. Thevenot, "Synthesis and Hot-pressing of Single-phase ZrCxOy and Two-phase ZrCxOy-ZrO2 Materials," *Int. J. High Technology Ceramics*, vol. 2, pp. 291–307, 1986.

[86] I. L. Shabalin, "Hafnium Monocarbide," in *Ultra-High Temperature Materials II*, Dordrecht: Springer Netherlands, 2019, pp. 145–248. doi: 10.1007/978-94-024-1302-1_3.

[87] H. Bittermann and P. Rogl, "Critical assessment and thermodynamic calculation of the binary system hafnium-carbon (Hf-C)," *Journal of Phase Equilibria*, vol. 18, no. 4, pp. 344–356, 1997, doi: 10.1007/s11669-997-0061-3.

[88] Z. Ren, "Review of HfCx and HfCx Based Composites since 1960s : Non-stoichiometric Characteristic, Powder Synthetic Methods, and Mechanical, Physical, Anti-oxidation Properties Improvements with Evolved Sintering Methods," 2020. doi: 10.13140/RG.2.2.24027.11045.

[89] A. L. Bowman, "The variation of lattice parameter with carbon content of tantalum carbide," *Journal of Physical Chemistry*, vol. 65, no. 9, pp. 1596–1598, 1961, doi: 10.1021/j100905a028.

[90] I. M. Pohrelyuk, V. M. Fedirko, and O. V. Tkachuk, "Effect of the rarefaction of an oxygen-containing medium on the formation of titanium oxynitrides," *Materials Science*, vol. 44, no. 1, pp. 64–69, 2008, doi: 10.1007/s11003-008-9044-8.

[91] K. Constant, R. Kieffer, and P. Ettmayer, "Über das pseudoternäre System 'ZrO'-ZrN-ZrC," *Monatsh Chem*, vol. 106, no. 4, pp. 823–832, 1975, doi: 10.1007/BF00900860.

[92] K. Constant, R. Kieffer, and P. Ettmayer, "Über das pseudoternäre System 'HfO'-HfN-HfC," *Monatsh Chem*, vol. 106, no. 4, pp. 973–981, 1975, doi: 10.1007/BF00900876.

[93] N. K. Tolochko *et al.*, "Absorptance of powder materials suitable for laser sintering," *Rapid Prototyp J*, vol. 6, no. 3, pp. 155–161, 2000, doi: 10.1108/13552540010337029.

[94] T. Mühler, C. M. Gomes, J. Heinrich, and J. Günster, "Slurry-based additive manufacturing of ceramics," *Int J Appl Ceram Technol*, vol. 12, no. 1, pp. 18–25, 2015, doi: 10.1111/ijac.12113.

[95] A. S. Bolokang, "Structural and Gas Sensing Properties of TiO2-Based (Sn, Mg) Nano Structures Induced by Mechanical Milling and Annealing," University of The Western Cape, 2013.

[96] R. Wróbel and W. Arabczyk, "Solid-gas reaction with adsorption as the rate limiting step," *Journal of Physical Chemistry A*, vol. 110, no. 29, pp. 9219–9224, 2006, doi: 10.1021/jp061947b.

[97] L. Fedunik-Hofman, A. Bayon, and S. W. Donne, "Kinetics of solid-gas reactions and their application to carbonate looping systems," *Energies (Basel)*, vol. 12, no. 15, 2019, doi: 10.3390/en12152981.

[98] M. Pijolat and L. Favergeon, "Kinetics and Mechanisms of Solid-Gas Reactions," vol. 6, 2018, pp. 173–212. doi: 10.1016/B978-0-444-64062-8.00011-5.

[99] H. Rode and V. Hlavacek, "Detailed Kinetics of Titanium Nitride Synthesis," vol. 41, no. 2, pp. 377–388, 1995.

[100] B. Khoshandam and R. V Kumar, "Producing Chromium Carbide Using Reduction of Chromium Oxide with Methane," *AIChE Journal*, vol. 52, no. 3, pp. 1094–1102, 2006, doi: 10.1002/aic.10712.

[101] X. He, Y. Zhu, A. Epstein, and Y. Mo, "Statistical variances of diffusional properties from ab initio molecular dynamics simulations," *NPJ Comput Mater*, vol. 4, no. 1, p. 18, 2018, doi: 10.1038/s41524-018-0074-y.

[102] J. C. Phys, L. Scotti, and A. Mottura, "Interstitial diffusion of O , N , and C in α -Ti from first-principles : Analytical model and kinetic Monte Carlo simulations," vol. 084701, no. November 2015, pp. 0–9, 2016, doi: 10.1063/1.4942030.

[103] B. W. Manley, "Comparative sinterability of combustion synthesized and commercial titanium carbide," United States, 1984.

[104] A. V. Gorshkov, "Empirical relation for the activation energy of diffusion in elemental substances," *Inorganic Materials*, vol. 36, no. 1, pp. 22–23, 2000, doi: 10.1007/BF02758373.

[105] A. B. Peters, "Additive Manufacturing of Complex Carbide Ceramics Using a Two-Step Powder Bed Fusion and Reaction Synthesis Approaches," *Manuscript in preparation*, 2022.

[106] I. Y. Guzman, "Reaction sintering in the technology of ceramics and refractories," *Glass and Ceramics*, vol. 42, pp. 273–277, 1985.

[107] A. Atkinson, A. J. Moulson, and E. W. Roberts, "Nitridation of high-purity silicon," *J Mater Sci*, vol. 10, no. 7, pp. 1242–1243, 1975, doi: 10.1007/BF00541409.

[108] A. B. Peters *et al.*, "A Reaction Synthesis Approach to Additively Manufacture Net-Shape Chromium Carbide and Non-Oxide Refractory Ceramics," *Addit Manuf*, vol. 34, Aug. 2020, doi: 10.1016/j.addma.2020.101186.

[109] M. Venkatraman and J. P. Neumann, "The C-Cr ( Carbon-Chromium ) System," *Bulletin of Alloy Phase Diagrams*, vol. 11, no. 2, pp. 152–159, 1990.

[110] M. N. Rahaman, *Ceramic Processing and Sintering*. Taylor & Francis, 2007.

[111] H. Song, R. L. Coble, and R. J. Brook, "The Applicability of Herring's Scaling Law to the Sintering of Powders BT - Materials Science Research: Volume 16 Sintering and Heterogeneous Catalysis," G. C. Kuczynski, A. E. Miller, and G. A. Sargent, Eds. Boston, MA: Springer US, 1984, pp. 63–79. doi: 10.1007/978-1-4613-2761-5_5.

[112] J. C. Maya and F. Chejne, "Novel model for non catalytic solid – gas reactions with structural changes by chemical reaction and sintering," *Chem Eng Sci*, vol. 142, pp. 258–268, 2016, doi: 10.1016/j.ces.2015.11.036.